\documentclass[prd,amsmath,preprintnumbers,superscriptaddress,floatfix,twocolumn,letterpaper,showpacs]{revtex4}
\usepackage{graphicx,placeins}
\usepackage{epsfig}
\usepackage{epstopdf}
\usepackage{wrapfig}
\usepackage{subfigure}
\usepackage{color}
\usepackage[utf8]{inputenc}
\usepackage{hyperref}
\usepackage{diagbox}
\usepackage{multirow}
\usepackage{booktabs}

\def\fm{\mathrm{fm}}
\def\mev{\mathrm{MeV}}

\def\lsim{\raise0.3ex\hbox{$<$\kern-0.75em\raise-1.1ex\hbox{$\sim$}}}
\def\gsim{\raise0.3ex\hbox{$>$\kern-0.75em\raise-1.1ex\hbox{$\sim$}}}

\newcommand{\bee}{\begin{equation}}
\newcommand{\ee}{\end{equation}}

\begin{document}
\preprint{}
%Title of paper
\title{Glueball spectrum from $N_f=2$ lattice QCD study on anisotropic lattices}

%\collaboration{CLQCD collaboration}

\author{Wei Sun}
\email{sunw@ihep.ac.cn}
\affiliation{Institute of High Energy Physics, Chinese Academy of Sciences, Beijing 100049, P.R. China}
\affiliation{School of Physics, University of Chinese Academy of Sciences, Beijing 100049, P.R. China}
\author{ Long-Cheng Gui}
\affiliation{Department of Physics and Synergetic Innovation Center for Quantum Effects and Applications,\\
	Hunan Normal University, Changsha 410081, P.R. China }
\affiliation{Key Laboratory of Low-Dimensional Quantum Structures and Quantum Control of Ministry of Education, Changsha 410081, P.R. China}
\author{Ying Chen}
\affiliation{Institute of High Energy Physics, Chinese Academy of Sciences, Beijing 100049, P.R. China}
\affiliation{School of Physics, University of Chinese Academy of Sciences, Beijing 100049, P.R. China}
\author{ Ming Gong}
\affiliation{Institute of High Energy Physics, Chinese Academy of Sciences, Beijing 100049, P.R. China}
\author{ Chuan Liu}
\affiliation{Collaborative Innovation Center of Quantum Matter, Peking University, Beijing 100871, P.R. China}
\affiliation{School of Physics and Center for High Energy Physics, Peking University, Beijing 100871, P.R. China}
\author{ Yu-Bin Liu}
\affiliation{School of Physics, Nankai University, Tianjin 300071, P.R. China}
\author{ Zhaofeng Liu}
\affiliation{Institute of High Energy Physics, Chinese Academy of Sciences, Beijing 100049, P.R. China}
\affiliation{School of Physics, University of Chinese Academy of Sciences, Beijing 100049, P.R. China}
\author{ Jian-Ping Ma}
\affiliation{Institute of Theoretical Physics, Chinese Academy of Sciences, Beijing 100080, P.R. China}
\author{ Jian-Bo Zhang}
\affiliation{Department of Physics, Zhejiang University, Hangzhou, Zhejiang 310027, P.R. China}

\collaboration{CLQCD Collaboration}
\noaffiliation

\begin{abstract}

 The lowest-lying glueballs are investigated in lattice QCD using $N_f=2$ clover Wilson fermion on anisotropic lattices.
 We simulate at two different and relatively heavy quark masses, corresponding to
  physical pion mass of $m_\pi\sim 938$ MeV and $650$ MeV. The quark mass dependence of the glueball masses
  have not been investigated in the present study.
 Only the gluonic operators built from Wilson loops are utilized in calculating the corresponding correlation functions. In the tensor channel, we obtain the ground state mass to be 2.363(39) GeV and 2.384(67) GeV at $m_\pi\sim 938$ MeV and $650$ MeV, respectively. In the pseudoscalar channel, when using the gluonic operator whose continuum limit has the form of $\epsilon_{ijk}TrB_iD_jB_k$, we obtain the ground state mass to be 2.573(55) GeV and 2.585(65) GeV at the two pion masses. These results are compatible with the corresponding results in the quenched approximation. In contrast, if we use the topological charge density as field operators for the pseudoscalar, the masses of the lowest state are much lighter (around 1GeV) and compatible with the expected masses of the flavor singlet $q\bar{q}$ meson. This indicates that the operator $\epsilon_{ijk}TrB_iD_jB_k$ and the topological charge density couple rather differently to the glueball states and $q\bar{q}$ mesons. The observation of the light flavor singlet pseudoscalar meson can be viewed as the manifestation of effects of dynamical quarks. In the scalar channel, the ground state masses extracted from the correlation functions of gluonic operators are determined to be around 1.4-1.5 GeV, which is close to the ground state masses from the correlation functions of the quark bilinear operators. In all cases, the mixing between glueballs and conventional mesons remains to be further clarified in the future.
\end{abstract}
\pacs{12.38.Gc, 14.40.Rt}
\maketitle

\section{Introduction}

Due to the self-interactions among gluons, Quantum Chromodynamics (QCD) admits the existence of a
new type of hadrons made up of gluons, usually called glueballs. Glueballs are of great physical
interests since they are distinct from the conventional $q\bar{q}$ mesons described in the constituent
quark model. Glueballs have been intensively studied by lattice QCD and other
theoretical methods~\cite{JAFFE1976201,CORNWALL1983431,Hou:1982dy,Brower:2000rp,
	Szczepaniak:2003mr,Narison:2005wc,Sanchis-Alepuz:2015hma}, 
for more details of this subject, see reviews in~\cite{Klempt:2007cp,Mathieu:2008me,Crede:2008vw,Ochs:2013gi}. Early lattice QCD studies in the quenched approximation show that the lowest pure gauge glueballs are the scalar, the tensor, and the pseudoscalar glueballs, with masses of 1.5-1.7 GeV, 2.2-2.4 GeV, and 2.6 GeV, respectively~\cite{Morningstar:1997,Morningstar:1999rf,Chen:2005mg}.

Experimentally,
there are several candidates for the scalar glueball, such as
$f_0(1370),f_0(1500),f_0(1710)$, however, none of them has been unambiguously identified as a glueball
state. On the other hand, $J/\psi$ radiative decays are usually regarded as an ideal hunting ground
for glueballs. A few lattice studies have been devoted to the calculation of the radiative
production rate of the pure scalar and tensor glueballs in the quenched
approximation~\cite{Gui:2012gx,Yang:2013xba}. The predicted production
rate of the scalar glueball is consistent with that of $f_0(1710)$, 
and supports $f_0(1710)$ to be either a good candidate
for the scalar glueball or dominated by a glueball component. 
The predicted production rate of the tensor glueball is roughly 1\%.
It is interesting to note that the BESIII Collaboration 
find that the tensor meson $f_2(2340)$ has large branching fractions
in the processes $J/\psi\rightarrow \gamma \eta\eta$~\cite{Ablikim:2013hq} and $J/\psi\rightarrow \gamma
\phi\phi$~\cite{Ablikim:2016hlu}.

Even though the quenched lattice QCD studies have provide some information on the existence
of glueballs, it is highly desired that full lattice QCD studies can be performed in the glueball sector. For the masses of the scalar and tensor glueballs, some preliminary unquenched lattice studies have given compatible results~\cite{Bali:2000vr,Hart:2001fp,Richards:2010ck,Gregory:2012hu}.
However, for the mass of the pseudoscalar glueball, a consensus has not been reached. For example, in Ref.~\cite{Richards:2010ck} the authors observed a pseudoscalar glueballs state
with a mass close to the result in the quenched approximation, but this is not confirmed by Ref.~\cite{Gregory:2012hu}. On the other hand, owing to the $U_A(1)$ anomaly, in the pseudoscalar channel, gluons can couple strongly to the flavor singlet pseudoscalar meson ($\eta'$ in the $N_f=2+1$ case) in the presence of dynamical quarks. Therefore, it is mandatory to identify the contribution of
the $\eta'$ meson before one draws any conclusions on the pseudoscalar glueball.
	
In this work, we attempt to investigate the glueball spectrum using the $N_f=2$ clover Wilson fermion gauge field configurations that we generated on anisotropic lattices. In order to check the quark mass dependence, we have generated two guage configuration ensembles with two different bare quark mass parameters which correspond to the physical pion masses $m_\pi\sim 600$ and $938$ MeV, respectively. The advantage of using anisotropic lattice is two-folds: on
the one hand, a large statistics can be obtained by a relatively low cost of
computational resources, on the other hand, the finer lattice spacing in the temporal direction can provide a better resolution for the signals of the desired physical states.
As the first step, we will focus on the lowest-lying glueball states, such as the scalar, the tensor and the pseudoscalar states. Secondly, we will pay more attention to the pseudoscalar channel. A recent $N_f=2+1$ lattice study showed that $\eta'$ could be probed by the topological charge density operator~\cite{Fukaya:2015ara}. In contrast, a similar study in the quenched approximation found a pseudoscalar with a mass compatible with that in the pure gauge theory~\cite{Chowdhury:2014mra}. Motivated by this, we use conventional Wilson loop operators to study lowest pseudoscalar glueball state and check for the lowest flavor singlet meson state with topological charge density operator on the same gauge ensembles.

This paper is organized as follows: Section~\ref{sec:lattice_setup} contains
a brief description for the generation of gauge field configurations.
Section~\ref{sec:details} presents the calculation details and the results of the glueball
spectrum. The study of the pseudoscalar channel using the topological charge density operator will
be discussed in Section~\ref{sec:pseudoscalar}, where we will also analyze the difference of the topological charge density
operator from the conventional gluonic operators for the pseudoscalar glueball in previous
quenched studies. Finally, we will give a summary and an outlook in Section~\ref{sec:summary}.

\begin{table}
\caption{\label{conf details} Parameters of configurations. The spatial lattice spacing $a_s$ is set by the calculation of the static potentials and the Sommer's scale parameter $r_0^{-1}=410(20)$ MeV. We also give the value of $a_t^{-1}$ in the physical units. }
\begin{center}
\begin{tabular}{ccccccc}
\hline
\hline
   $\beta$      &   $m_0$  &   $L^3\times T$    &   $\xi$  &   $a_s$(fm)   &  $a_t^{-1}$ (GeV)    &    $N_{conf}$\\\hline
   $2.5$        &  0.05   &    $12^3\times128$  &    5     &   0.114(1)    &  8.654(76)                     &  4800  \\
   $2.5$        &  0.06   &    $12^3\times128$  &    5     &   0.118(1)    &  8.360(76)                     &  10400 \\\hline
\hline
\end{tabular}
\end{center}
\end{table}
\section{Lattice setup}
\label{sec:lattice_setup}
The gauge action we used is the tadpole improved gluonic action on anisotropic lattices~\cite{Morningstar:1997}:
\begin{eqnarray}\label{key}
S_g = &-& \beta \sum_{i>j} \left[ \frac{5}{9} \frac{Tr P_{ij}}{\gamma_g u_s^4} - \frac{1}{36}\frac{Tr R_{ij}}{\gamma_g u_s^6}-\frac{1}{36}\frac{Tr R_{ji}}{\gamma_g u_s^6} \right] \nonumber\\
	&-& \beta \sum_i \left[ \frac{4}{9}\frac{\gamma_g Tr       P_{0i}}{u_s^2} - \frac{1}{36}\frac{\gamma_g Tr R_{i0}}{u_s^4} \right]
\end{eqnarray}
where $P_{ij}$ is the usual plaquette variable and $R_{ij}$ is the $2 \times 1$ Wilson loop on the lattice. The parameter $u_s$, which we take to be the forth root of the average spatial plaquette value, incorporates the usual tadpole improvement and $\gamma_g$ designates the gauge aspect bare ratio of the anisotropic lattice, denoted as $\xi_0$ in our former quenched studies~\cite{Su:2004sc}. %arxiv:0412034
Although $\gamma_g$ suffers only small renormalization with the tadpole improvement \cite{PhysRevD.48.2250}, we have to tune it by determining the renormalized anisotropy ratio $\xi_g$. As for the tadpole improvement parameter $u_t$ for temporal gauge links, we take the approximation
$u_t\approx 1$ following the conventional treatment of the anisotropic lattice setup.

We use the Wilson-loop ratios approach, with which the finite volume artifacts mostly cancel \cite{Umeda:2003pj,Klassen:1998ua}. We measure the ratios
\begin{eqnarray}
R_{ss}(x,y) &=& \frac{W_{ss}(x,y)}{W_{ss}(x+1,y)} \to e^{-a_sV_s(ya_s)},\\
R_{st}(x,t) &=& \frac{W_{st}(x,t)}{W_{ss}(x+1,t)} \to e^{-a_sV_s(ta_t)}
\label{eq:wlsnlp_ratio}
\end{eqnarray}
and expect the spatial and temporal behaviors being the same at the correct $\xi_g$.

Therefore we find $\xi_g$ by minimizing
\begin{equation}
L(\xi_g) = \sum_{x,y} \frac{(R_{ss}(x,y)-R_{st}(x,\xi_g y))^2}{(\Delta R_s)^2+(\Delta R_t)^2}
\label{eq:gauge_anisotroy}
\end{equation}
where $\Delta R_s$ and $\Delta R_t$ are the statistical errors of $R_{ss}$ and $R_{st}$.
We interpolate $R_{st}(x,\xi_g y)$ and its error with a cubic spline interpolation at non-integer $\xi_g y$. Since small $x,y$ may introduce short-range lattice effects and large ones contribute only fluctuations, we scan and test different ranges and finally choose $x,y \in \{2,3,4,5\}$.

We adopt the anisotropic clover fermion action in the fermion sector \cite{Edwards:2008ja}:
\begin{eqnarray}
S_f &=& \sum_x \bar\psi(x) \left[ m_0
+ \gamma_t \hat W_t
+ \sum_s  \frac{1}{\gamma_f} \gamma_s \hat W_s \right. \nonumber\\
  & &\left. - \frac{1}{4 u_s^2}\left( \frac{\gamma_g}{\gamma_f}
+ \frac{1}{\xi} \right) \sum_s \sigma_{ts} \hat F_{ts} \right.\nonumber\\
  & &\left. + \frac{1}{u_s^3} \frac{1}{\gamma_f} \sum_{s<s^\prime} \sigma_{ss^\prime} \hat F_{ss^\prime}
\right] \psi(x)
\end{eqnarray}
where $\hat F_{\mu \nu} = \frac{1}{4} {\rm Im}(P_{\mu\nu}(x))$ and the dimensionless Wilson operator reads
\begin{eqnarray}
\hat W_\mu &=& \nabla_\mu -\frac{1}{2} \gamma_\mu \Delta_\mu \nonumber \\
\nabla_\mu f(x) &=& \frac{1}{2} \left[ U_\mu(x)f(x+\mu)-U_\mu^\dag(x-\mu)f(x-\mu)\right] \nonumber \\
\Delta_\mu f(x) &=& U_\mu(x)f(x+\mu)+U_\mu^\dag(x-\mu)f(x-\mu)-2f(x). \nonumber
\end{eqnarray}
The bare fermion aspect ratio $\gamma_f$ is also tuned to make sure that the measured aspect ratio $\xi_f \approx \xi_g \approx \xi=5$.
$\xi_f$ is measured from the dispersion relation of the pseudoscalar and vector mesons
\begin{equation}
E(p)^2a_t^2 = m^2a_t^2 + \frac{|\vec p|^2a_s^2}{\xi_f^2}.
\label{eq:dispersion_relation}
\end{equation}
where $\vec p=2\pi \vec k/L_s$ is the momentum on the lattice with periodic spatial boundary conditions.

We generate two gauge ensembles on the $12^3\times128$ anisotropic lattice at $\beta=2.5$ with the bare quark mass parameters $m_0=0.05$ and $m_0=0.06$. The lattice spacings $a_s$ are set by calculating the static potential parameterized as $V(r)=V_0+\alpha/r+\sigma r$.
Using the Sommer scale parameter $r_0^{-1}=410(20)$ MeV defined through $r^2\frac{d V(r)}{d r}|_{r=r_0} = 1.65$, we can
determine the ratio
\begin{equation}
\frac{r_0}{a_s} = \sqrt{\frac{1.65+\alpha}{\sigma a_s^2}}
\label{eq:r0_over_as}
\end{equation}
where $\alpha$ and $\sigma a_s^2$ are derived from the fit to calculated potential $V(r)=V(\hat{r}a_s)$ with $\hat{r}$ being the spatial distance in the lattice units. Finally, $a_s$ is inverted to the values in the physical units by the Sommer's scale parameter $r_0^{-1}=410(20)$ MeV. The ensemble parameters are listed in Table~\ref{conf details}, where we also give the physical values
of $a_t^{-1}$ for the two ensemble.

The pion masses on the two ensembles are measured to be $938$ MeV and $650$ MeV respectively.
In the following, we use these $m_\pi$'s to label the gauge ensembles for convenience.
Apart from the pion masses, we also calculate the masses of the vector meson and scalar meson for calibration, which are listed in Table~\ref{meson_spec}. We
use the conventional $I=1$ vector and scalar quark bilinear operators as sink operators and the corresponding Gaussian smeared wall source operators to
calculate the correlation functions. There is no ambiguity for the vector meson masses $m_V$'s since they are all below the two-pion threshold. For the scalar,
we actually deal with $a_0$ whose two-body strong decay mode is mainly $\eta'\pi$ (there is only one $I=0$ pseudoscalar meson for $N_f=2$, which is taken as the counterpart of the (approximately) flavor-singlet $\eta'$ in the $N_f=3$ case). At $m_\pi\sim 938$ MeV, the calculated mass in $a_0$ channel is $1473(28)$ MeV, which must be the mass of $a_0$ since it lies below two-pion threshold and certainly below the $\eta'\pi$ threshold. At $m_\pi\sim 650$ MeV, $m_{\eta'}$ is estimated to be $m_{\eta'}\sim 890$ MeV (see below in Sec. 4), thus the mass value of $1362(53)$ MeV is also below the $\eta'\pi$ threshold and can be taken
as the mass of $a_0$ scalar at this pion mass. In order to calculate the $I=0$ scalar meson mass, the disconnected diagrams (quark annihilation diagrams) should
be considered. We have not done this yet, but as a rough estimate, 
we take the $a_0$ mass as an approximation to the mass of the isoscalar scalar meson.
\addtolength{\tabcolsep}{15pt}
\begin{table*}[!htbp]
\caption{\label{meson_spec} The masses of the ground state pseudoscalar, vector and scalar mesons (these are actually isovector mesons since we ignore the disconnected contributions). The measured values are also inverted to the values in physical units through $a_t^{-1}$ in Table~\ref{conf details}}
\begin{center}
	\begin{tabular}{cc|cc|cc}
	\hline
	\hline
	$m_{PS} a_t$	  &	$m_{PS}(\mev)$	&	$m_Va_t$	&	$m_V(\mev)$	&	$m_Sa_t$	&	$m_S(\mev)$	\\\hline
	$0.07508(50)$ &	$650(4)$			&	$0.1147(19)$&	$993(16)$		&	$0.1574(61)$&	$1362(53)$		\\\hline
	$0.1119(4)$	  &	$938(3)$			&	$0.1388(12)$&	$1164(10)$		&	$0.1757(34)$&	$1473(28)$		\\\hline
	\hline
	\end{tabular}
\end{center}
\end{table*}
\addtolength{\tabcolsep}{-15pt}

\section{Numerical details}
\label{sec:details}

In this work, the spectrum of the lowest-lying glueballs in three specific channels, namely scalar, tensor and pseudoscalar will be explored. The interpolating operators for these states are
pure gluonic operators which have been extensively adopted in the previous quenched lattice studies.
In other words, in each specific channel, no operators involving quark fields are included.
This of course is only an approximation, assuming that the gluon-dominated state that we are after
can be well-described by gluonic operators. Needless to say, mixing with the quark operators should
be considered later on, especially for cases where the mixing is severe.
For completeness, we briefly recapitulate the major ingredients of glueball spectrum computation in
the following. One can resort to~\cite{Chen:2005mg} for further details.

\subsection{Variational method}

The continuum $SO(3)$ spatially rotational symmetry is broken into the discrete symmetry described
by the octahedral point group $O$ on the lattice, whose irreducible representations $R$ are
labeled as $A_1,A_2,E,T_1,T_2$, and have dimensions 1, 1, 2, 3, 3 respectively. Therefore, the
lattice interpolation fields for a glueball of $J^{PC}$ quantum number should be denoted by $R^{PC}$ with $R$
the irreducible representation of $O$ which may include the components of $J$ in the continuum
limit. The parity $P=\pm$ and the charge conjugation $C=\pm$ can be realized by considering the
transformation properties under the spatial reflection and time reversal operations. Since the
octahedral group $O$ is a subgroup of $SU(2)$, the subduced representation of $SU(2)$ with respect
to $O$ is reducible in general (for integer spin, this occurs for $J\geq 2$). Table~\ref{subduced rep} shows
the reduction of the subduced representation of $SU(2)$ up to $J=5$. For instance, the scalar and
pseudoscalar with $J=0$ states are represented by $A_1$, tensor states with $J=2$ are reduced to
direct sum of $E$ and $T_2$, i.e. $(J=2)\downarrow O=E\bigoplus T_2$.
\setlength{\arrayrulewidth}{0.6pt}
\addtolength{\tabcolsep}{2pt}
\begin{table}[!htbp]
\caption{\label{subduced rep} Reduction of subduced representation of $SU(2)$ with respect to octahedral group $O$ up to $J=5$}
\begin{center}
\begin{tabular}{|l|ccccccccccc|}
\hline
\diagbox{R}{J}  &   $0$     & \ \ \ \ &    $1$ & \ \ \ \ &  $2$ & \ \ \ \ &$3$   & \ \ \ \ &  $4$   & \ \ \ \ &  $5$ \\\hline
$A_1$           &   1       &   &0       &   &0  	&   &0     &   &1   	&   & 0    \\
$A_2$           &   0       &   &0       &   &0   	&   &1     &   &0   	&   & 0 \\
$E$             &   0       &   &0       &   &1   	&   &0     &   &1   	&   & 1 \\
$T_1$           &   0       &   &1       &   &0   	&   &1     &   &1   	&   & 2 \\
$T_2$           &   0       &   &0       &   &1   	&   &1     &   &1   	&   & 1 \\
\bottomrule
\hline
\end{tabular}
\end{center}
\end{table}
\addtolength{\tabcolsep}{2pt}
\setlength{\arrayrulewidth}{0.4pt}
\begin{figure}[!htbp]
\centerline{\includegraphics[width=6.0cm]{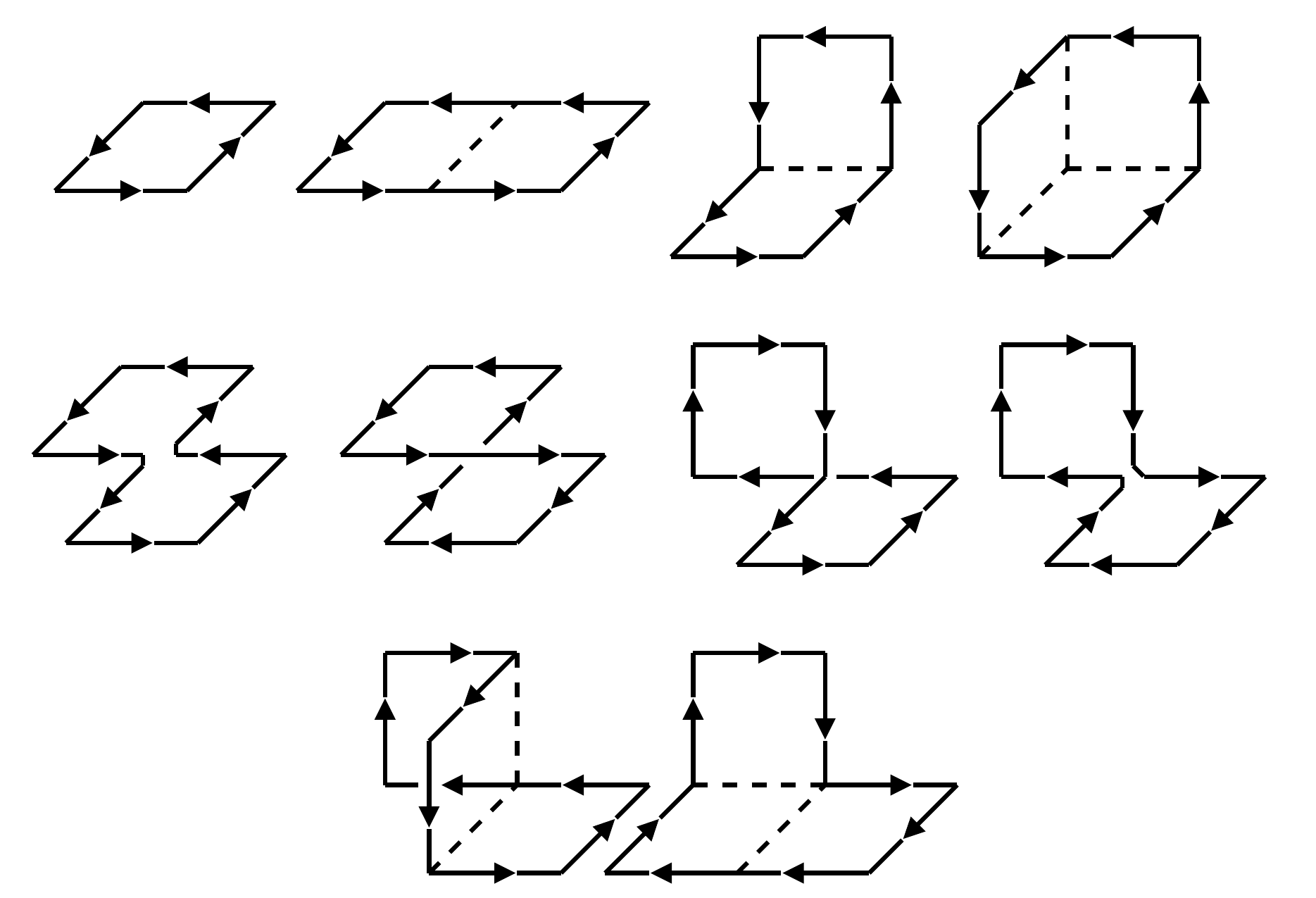}}
\caption{\label{loops} Prototypes of Wilson loops for the construction of glueball operators~\cite{Morningstar:1999rf,Chen:2005mg}.}
\end{figure}
As described in~\cite{Chen:2005mg}, we use Wilson loops (up to 8 gauge links) shown in Fig.~\ref{loops}. Each irrep $R$ of group $O$ can be realized by the specific linear combination of
its 24 copies of a prototype Wilson loop under the 24 rotation operations of $O$. The combination
coefficients of each $R$ can be found in~\cite{Chen:2005mg}. So each prototype may provide a
different realization of $R$. On the other hand, the Wilson loops mentioned above can be built from
smeared gauge links, such that different smearing schemes can provide more realizations of the gluonic operators. In practice, we have four different realization of each $R$ by
choosing different prototypes. For the smearing of gauge links, we adopt 6 smearing schemes by
combining the single-link and double-link smearing procedures with different iteration sequences.
Finally we have a set of 24 different gluonic operators, $\{\phi^{(R)}_\alpha, \alpha=1,2,\ldots,24\}$,  for each $R^{PC}$.

Based on these operator sets, we use the variational method to get the optimized operators
$\mathcal{O}^{(R)}$ which mostly project to specific glueball states. In each symmetry channel $R$,
we first calculate the $24\times 24$ correlation matrix $C^{(R)}(t)$,
\begin{equation}
\label{correMatrix} C^{(R)}_{\alpha\beta}(t)=\sum_\tau\langle
0|\bar{\phi}_\alpha^{(R)}(t+\tau)\bar{\phi}_\beta^{(R)\dagger}(\tau)|0\rangle,\quad\alpha,\beta=1,...,24
\end{equation}
where $\bar{\phi}^{(R)}_\alpha$ is the vacuum-subtracted operator of $\phi^{(R)}_\alpha$,
\begin{equation}
\label{vacuumSub}
\bar{\phi}_\alpha^{(R)}(t)=\phi_\alpha^{(R)}(t)-\langle0|\phi_\alpha^{(R)}(t)|0\rangle.
\end{equation}
In practice, we only apply the vacuum subtraction to the operators in $A_1^{++}$ channel.
Secondly, we solve the following generalized eigenvalue problem,
\begin{equation}
\label{GEVP} \mathbf{C}^{(R)}(t_0)\mathbf{v}_i^{(R)}=\lambda_i(t_0)
\mathbf{C}^{(R)}(0)\mathbf{v}_i^{(R)},
\end{equation}
where $\mathbf{v}_i^{(R)}$ is the $i$-th eigenvector, and $\lambda_i\equiv e^{-\bar{m}_i(t_0)t_0}$
is the $i$-th eigenvalue where $\bar{m}_i(t_0)$ is dependent on $t_0$ and is close to the energy of
the $i$-th state. For all the $R$ channels, we use $t_0=1$. It is expected that the eigenvector
$\mathbf{v}^{(R)}_i$ gives the linearly combinational coefficients of operators
$\bar{\phi}^{(R)}_\alpha$ to build an optimal operator $\Phi^{(R)}_i$ which overlaps mostly to the
$i$-th state,
\begin{equation}
\Phi^{(R)}_i(t)=\sum\limits_{\alpha=1}^{24} v^{(R)}_{i,\alpha}\bar{\phi}^{(R)}_\alpha(t).
\end{equation}

\subsection{Data analysis}

In this work, the correlation function of the optimal operator $\Phi^{(R)}_i$ for the $i$-th state
is calculated as
\begin{equation}
\label{optimal corre}
\tilde{C}_i^{(R)}(t)=\sum_\tau\langle 0|\Phi_i^{(R)}(t+\tau)\Phi_i^{(R)\dagger}(\tau)|0\rangle,
\end{equation}
where we do the summation over the temporal direction to increase the statistics. Accordingly, the
effective mass is defined as
\begin{equation}
m_{i,\rm eff}^{(R)}(t)=\ln\left(\frac{\tilde{C}_i^{(R)}(t)}{\tilde{C}_i^{(R)}(t+1)}\right).
\end{equation}
We divide the measurements
into bins with each bin including 100 measurements.
The statistical errors are obtained by the one-bin-eliminating jackknife
analysis.

For $A_1^{++}$ channel, the subtraction of the vacuum is very subtle. Even though we have $O(10^4)$ gauge configurations in each
ensemble, when we perform the jackknife analysis above after subtracting the vacuum expectation values of the operator, we find
there is still a residual (negative) constant term in the correlation function, which makes the effective mass $m_{i,\,{\rm eff}}(t)$
going upward when $t$ is large. This problem can be attributed to the large fluctuation of gauge configurations in the presence of sea quarks.
To circumvent this difficulty, we adopt a vacuum-subtraction scheme by subtracting the correlation function $C(t)$ with the shifted one $C(t+\delta t)$,
\begin{equation}
\label{shifted_corre}
\bar{C}_i^{A_1^{++}}(t)=\tilde{C}_i^{A_1^{++}}(t)-\tilde{C}_i^{A_1^{++}}(t+\delta t),
\end{equation}
whose spectral expression is
\begin{equation}
\bar{C}_i^{A_1^{++}}(t)=\sum\limits_j W_{ij}^{A_1^{++}}(1-e^{-m_j\delta t})e^{-m_j t}\equiv \sum\limits_j W_{ij}'^{A_1^{++}}e^{-m_j t},
\end{equation}
where $W_{ij}^{A_1^{++}}$ is the spectral weight of the $j$-th state in $\tilde{C}_i^{A_1^{++}}(t)$. Obviously, the possible constant term
cancels with the spectrum unchanged.
In practice, we take $\delta t=5 a_t$.

\begin{figure*}[ht]
\centering
\subfigure[$m_\pi\sim938~\mev$]{\label{a1pp:900}\includegraphics[width=6.5cm]{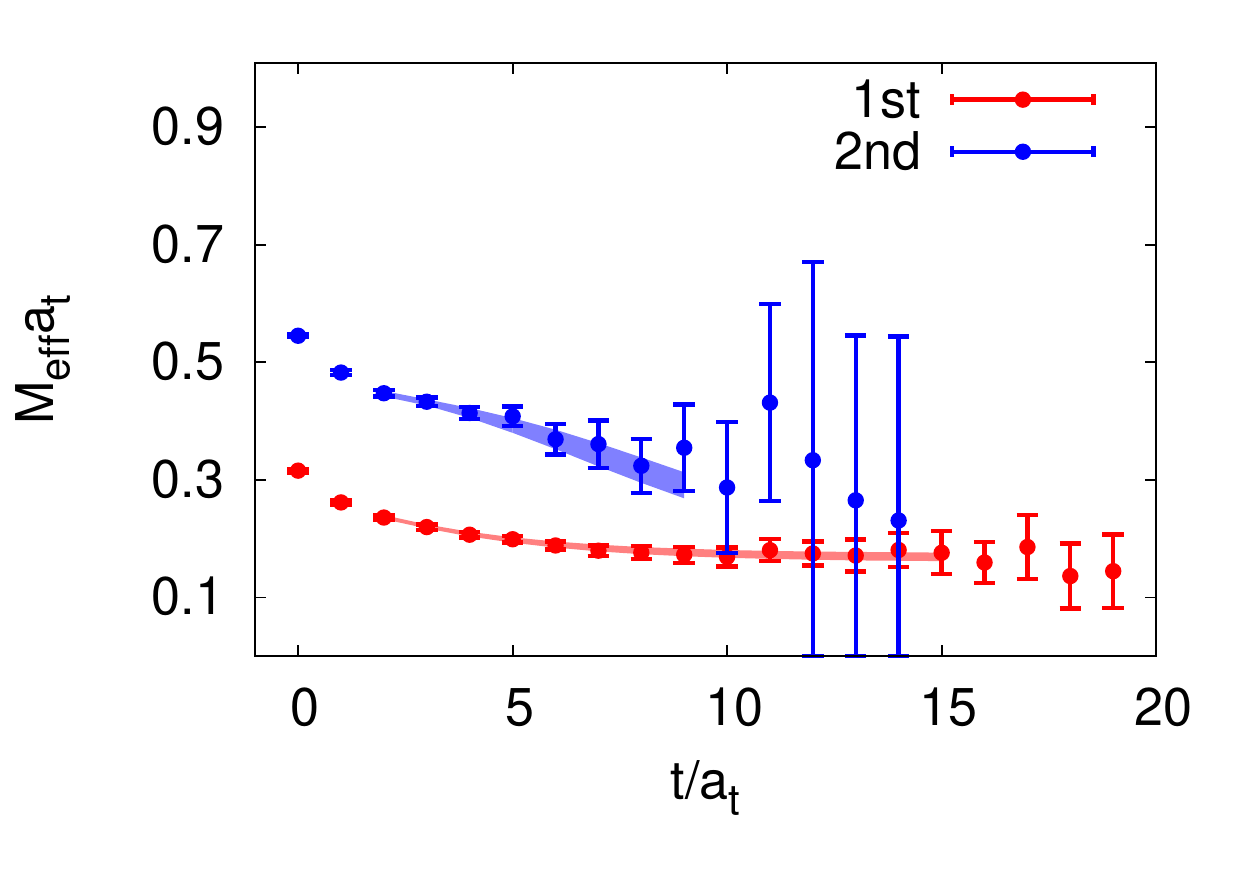}}\hfil
\subfigure[$m_\pi\sim650~\mev$]{\label{a1pp:600}\includegraphics[width=6.5cm]{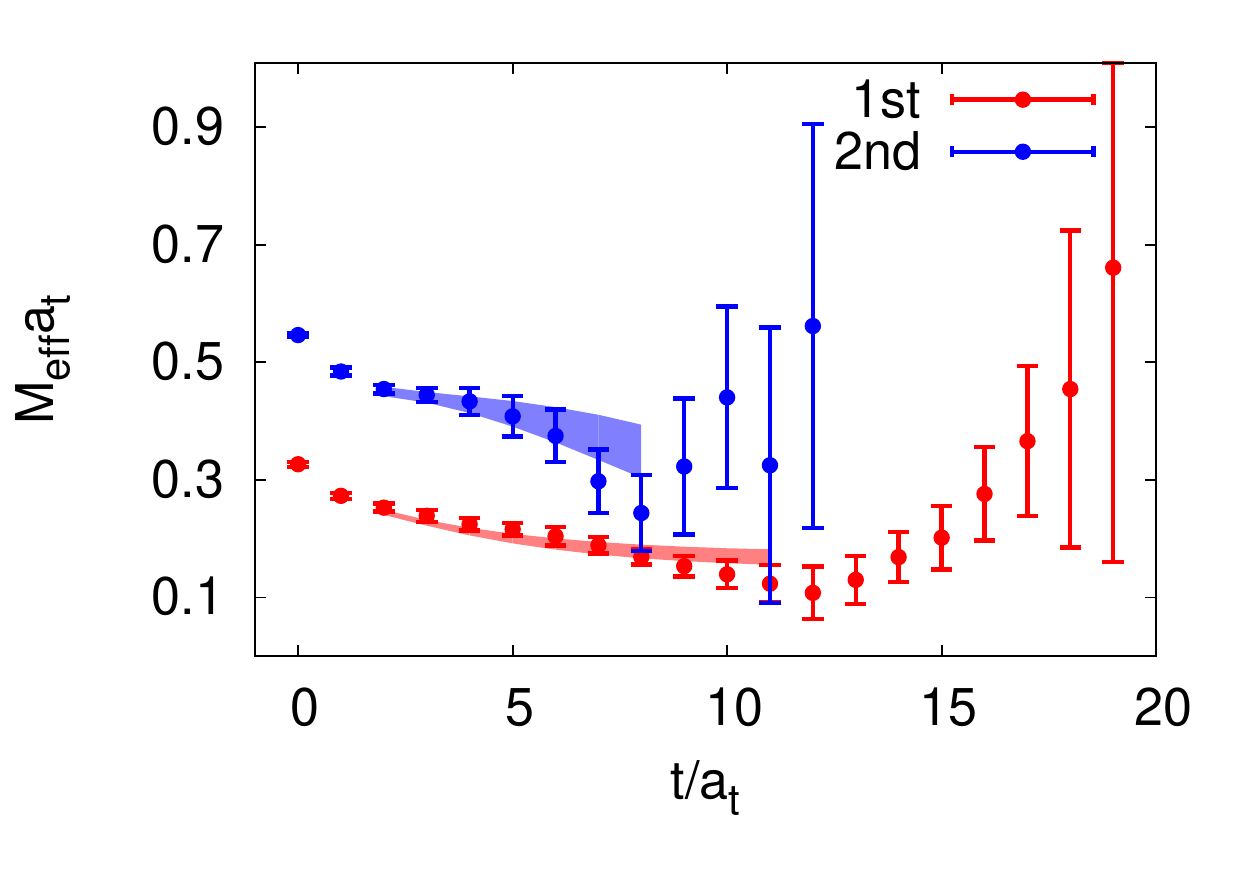}}
\caption{Effective mass plateaus of $\bar{C}_1^{(R)}(t)$ (red points) and $\bar{C}_2^{(R)}(t)$ (blue points) in the $R=A_1^{++}$ channel. The left and the right panel show the results at $m_\pi\sim 938~\mev$ and $m_\pi\sim650~\mev$, respectively.  The shaded bands are plotted with the best fit parameters using model of Eq.~(\ref{fit_model}) in the illustrated time range.} \label{a1pp}
\end{figure*}
\begin{figure*}[ht]
\centering
\subfigure[$m_\pi\sim938~\mev$]{\label{a1mp:900}\includegraphics[width=6.5cm]{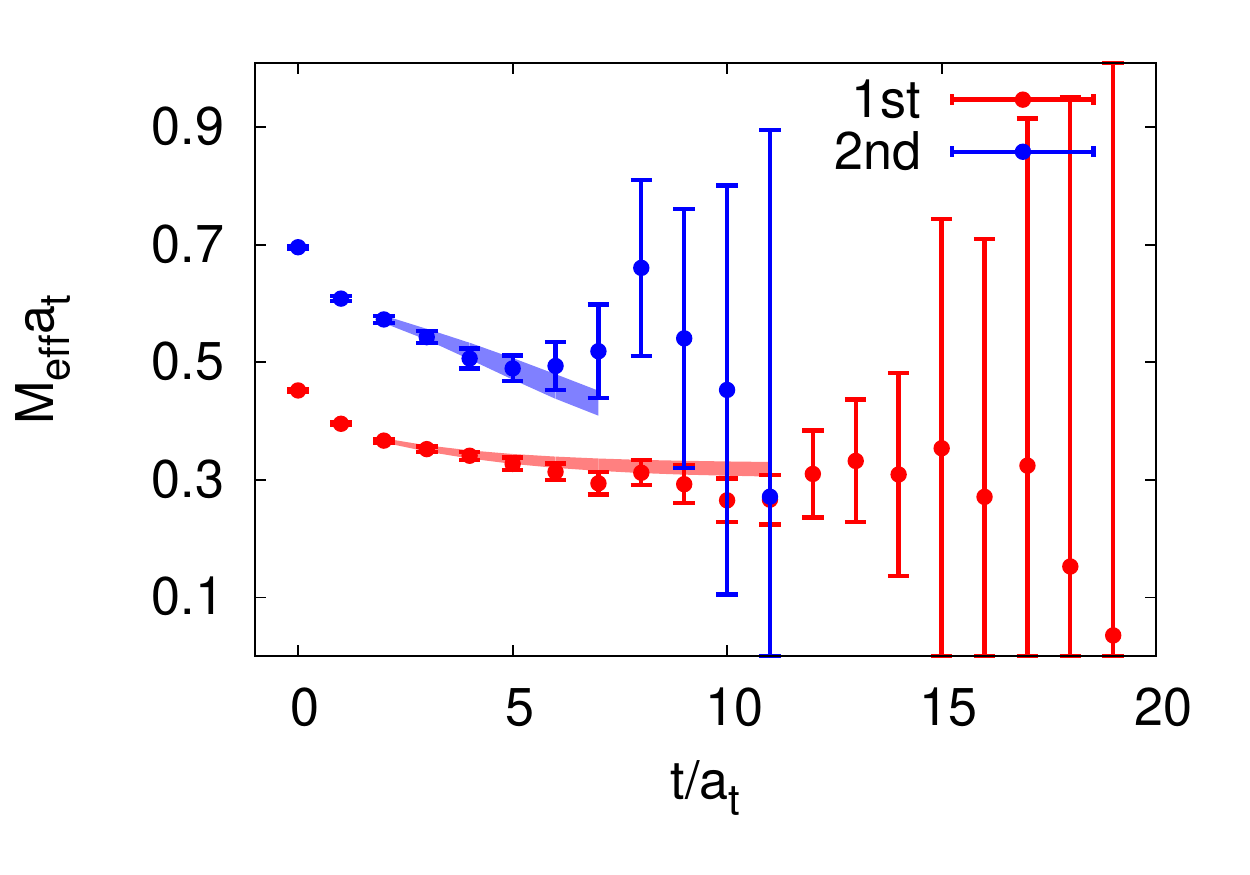}}\hfil
\subfigure[$m_\pi\sim650~\mev$]{\label{a1mp:600}\includegraphics[width=6.5cm]{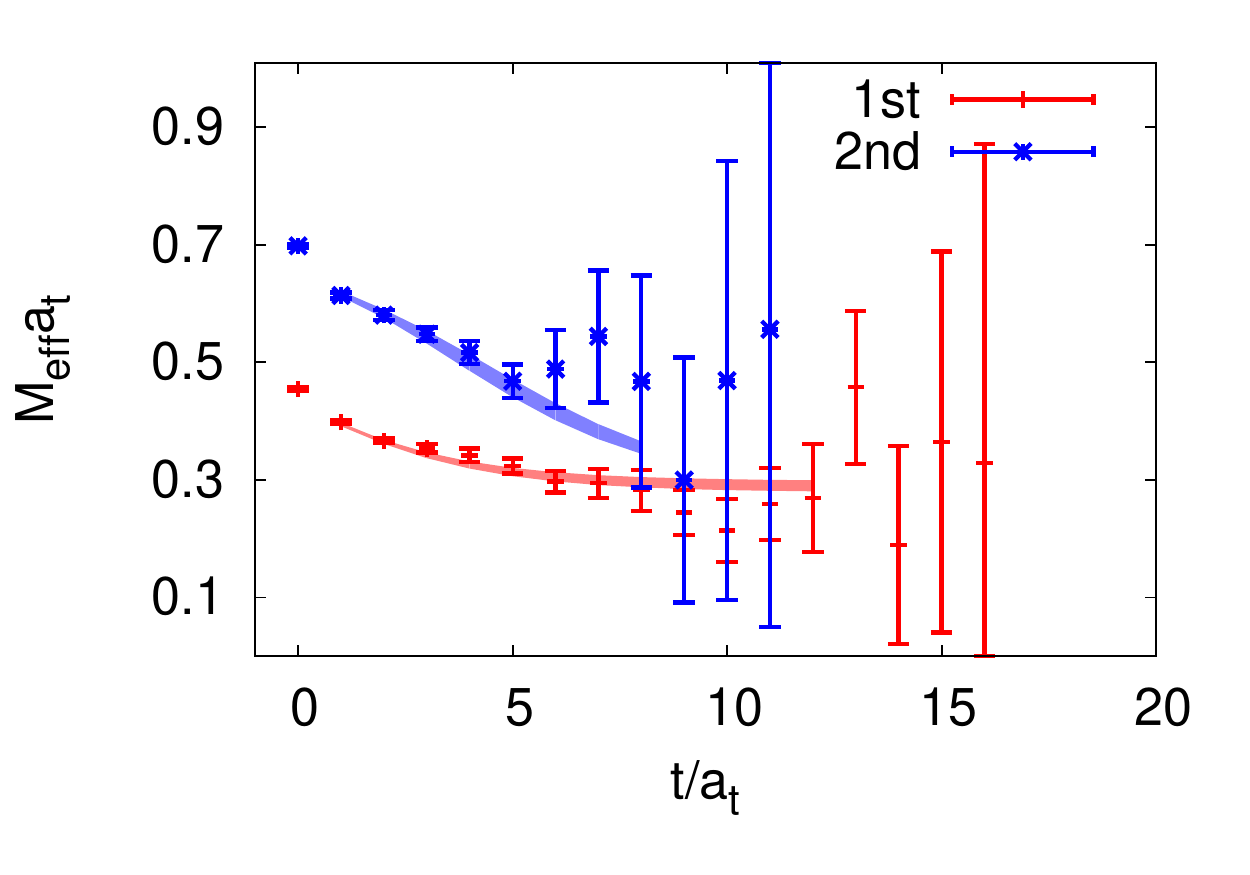}}
\caption{Effective mass plateaus of $\tilde{C}_1^{(R)}(t)$ (red points) and $\tilde{C}_2^{(R)}(t)$ (blue points) in the $R=A_1^{-+}$ channel. The left and the right panel show the results at $m_\pi\sim 938~\mev$ and $m_\pi\sim650~\mev$, respectively.  The shaded bands are plotted with the best fit parameters using model of Eq.~(\ref{fit_model}) in the illustrated time range. } \label{a1mp}
\end{figure*}
\begin{figure*}[ht]
\centering
\subfigure[$m_\pi\sim938~\mev$]{\includegraphics[width=6.5cm]{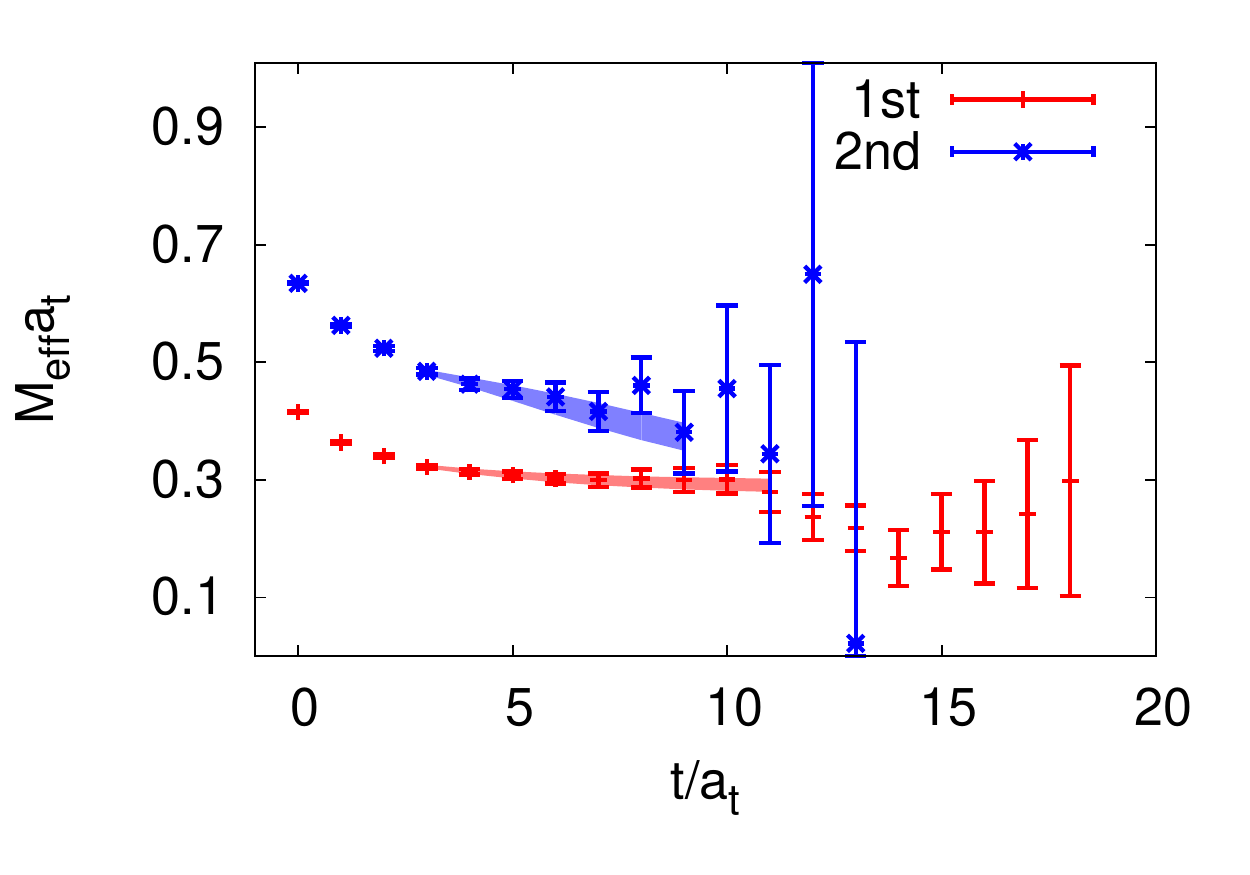}}\hfil
\subfigure[$m_\pi\sim650~\mev$]{\includegraphics[width=6.5cm]{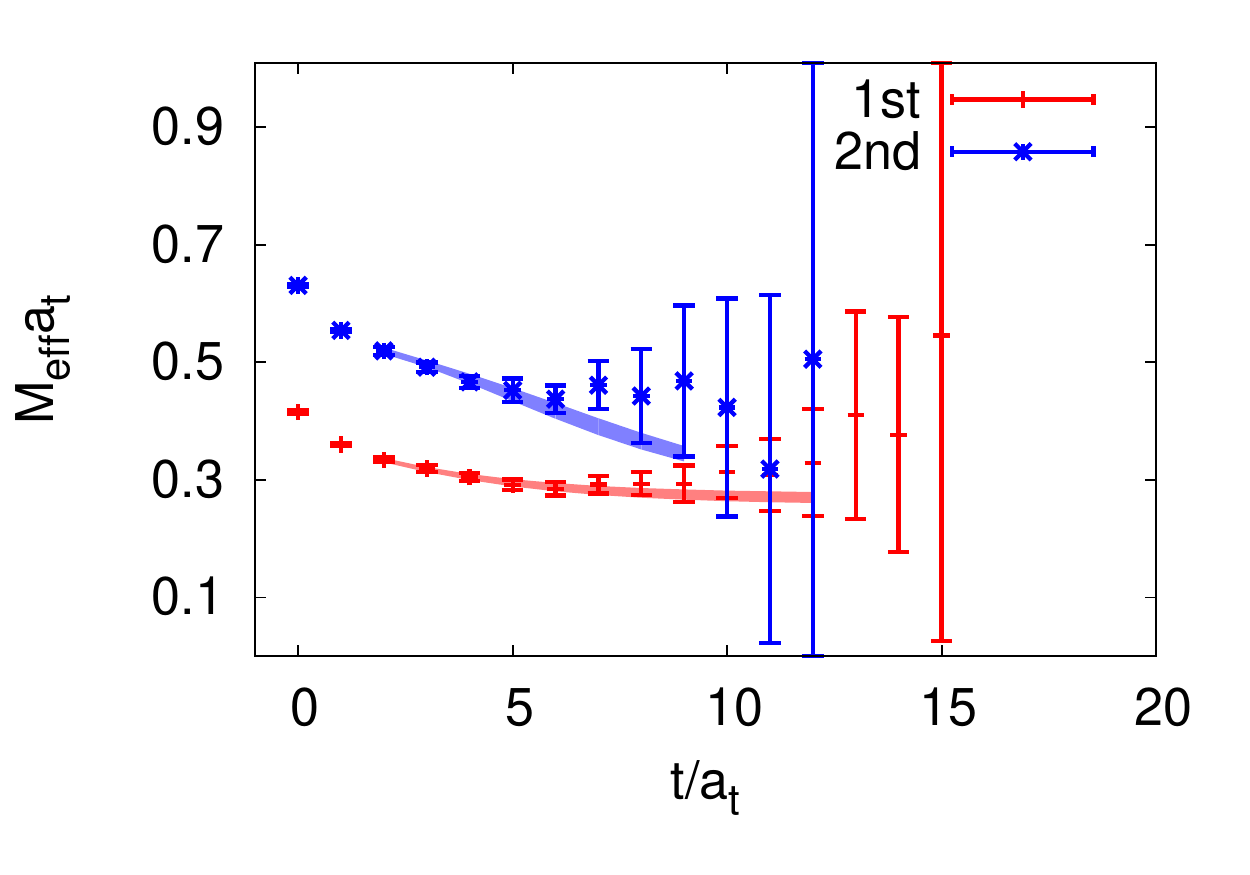}}
\caption{Effective mass plateaus of $\tilde{C}_1^{(R)}(t)$ (red points) and $\tilde{C}_2^{(R)}(t)$ (blue points) in the $R=E^{++}$ channel. The left and the right panel show the results at $m_\pi\sim 938~\mev$ and $m_\pi\sim650~\mev$, respectively.  The shaded bands are plotted with the best fit parameters using model of Eq.~(\ref{fit_model}) in the illustrated time range.} \label{epp}
\end{figure*}
\begin{figure*}[ht]
\centering
\subfigure[$m_\pi\sim938~\mev$]{\includegraphics[width=6.5cm]{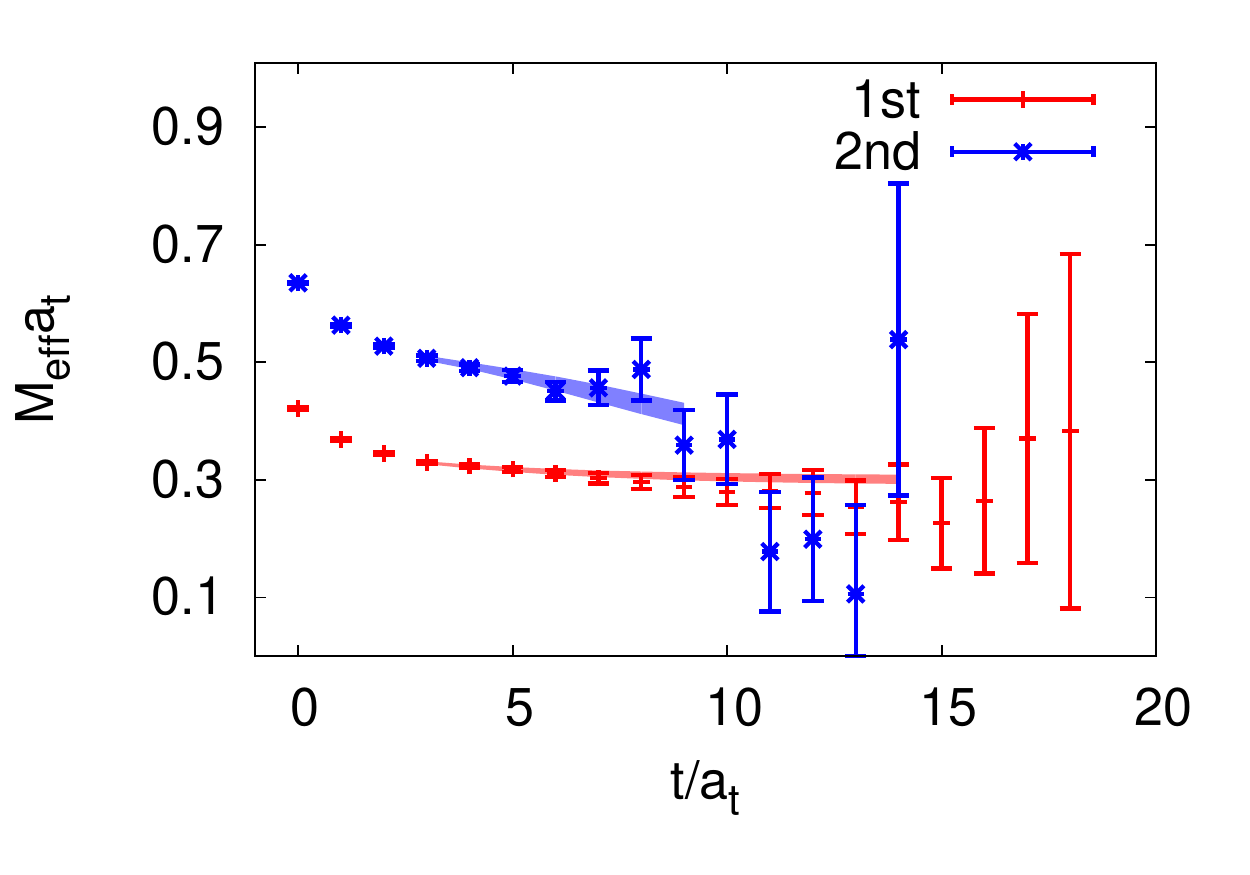}}\hfil
\subfigure[$m_\pi\sim650~\mev$]{\includegraphics[width=6.5cm]{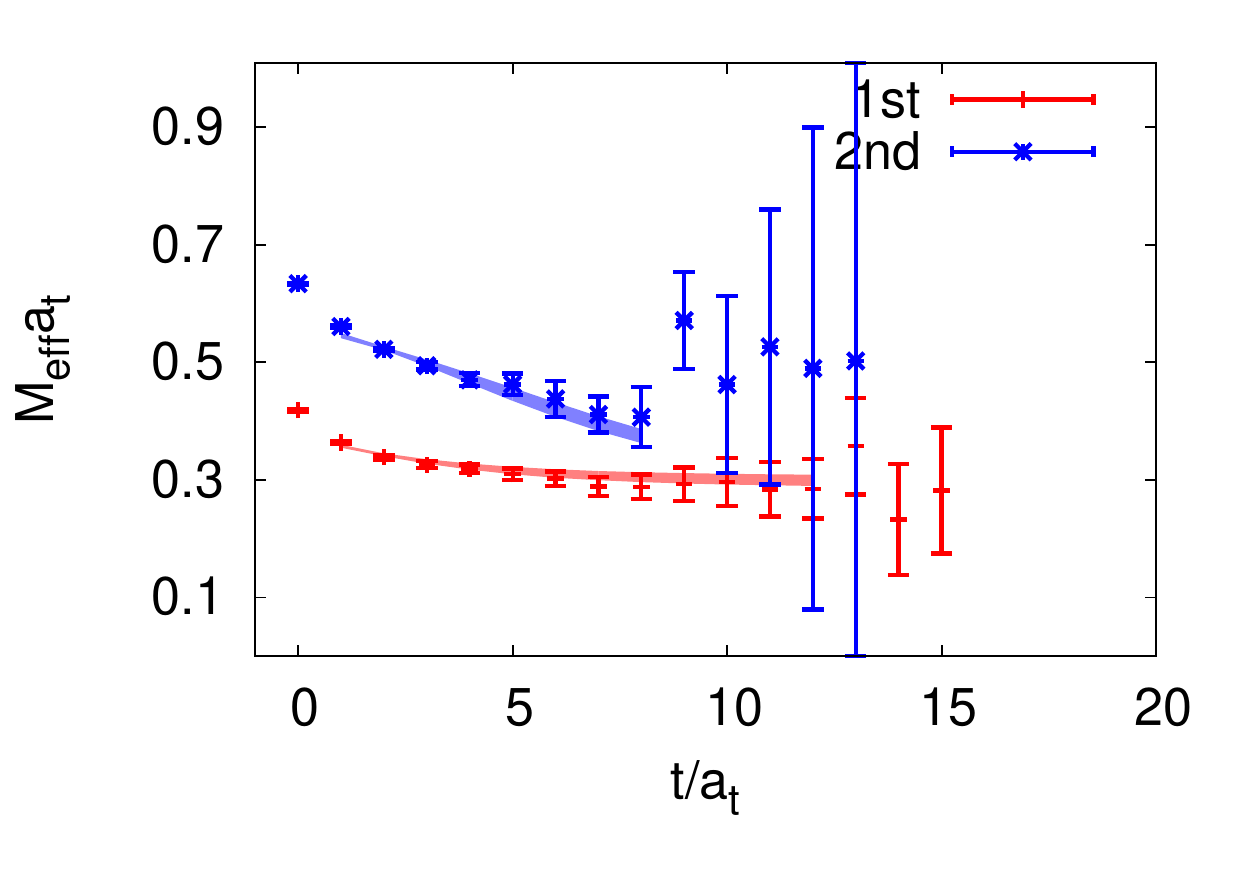}}
\caption{Effective mass plateaus of $\tilde{C}_1^{(R)}(t)$ (red points) and $\tilde{C}_2^{(R)}(t)$ (blue points) in the $R=T_2^{++}$ channel. The left and the right panel show the results at $m_\pi\sim 938~\mev$ and $m_\pi\sim650~\mev$, respectively.  The shaded bands are plotted with the best fit parameters using model of Eq.~(\ref{fit_model}) in the illustrated time range. } \label{t2pp}
\end{figure*}

We focus on the $R^{PC}=A_1^{++},A_1^{-+},E^{++}$, and $T_2^{++}$ channels in this work. For all these channels, the effective masses of
$\tilde{C}_1^{R}(t)$ (red points) and $\tilde{C}_2^{R}(t)$ (blue points) ($\bar{C}_i^{A_1^{++}}$ for  $A_1^{++}$ channel) are plotted in Fig.~\ref{a1pp},~\ref{a1mp},~\ref{epp} and~\ref{t2pp}, respectively.
In each figure, the left panel shows the result at $m_\pi\sim938~\mev$, and the right panel is for $m_\pi\sim650~\mev$.
Even though we have a set of 24 operators for each channel, it is seen that the effective masses do not show plateaus from the very early time slices. This is very different from the case in the quenched approximation. One important reason for this is that, in each channel, the spectrum of the full QCD is much more complicated than in the quenched approximation due to the sea quarks. This is true in principle, since $q\bar{q}$ states and multi-hadron states with the same quantum number do contribute to the corresponding correlation function in the presence of sea quarks.
%\begin{figure*}[!htbp]
%	\centering
%	\subfigure[$m_\pi\sim938~\mev$]{\includegraphics[width=7.0cm]{figs/t2++_104.pdf}}
%	\subfigure[$m_\pi\sim650~\mev$]{\includegraphics[width=7.0cm]{figs/t2++_48.pdf}}
%	\caption{Effective mass plateaus of $\tilde{C}_1^{(R)}(t)$ (red points) and $\tilde{C}_2^{(R)}(t)$ (blue points) in the $R=T_2^{++}$ channel. The left and the right panel show the results at $m_\pi\sim 938~\mev$ and $m_\pi\sim650~\mev$, respectively.  The shaded bands are plotted with the best fit parameters using model of Eq.~(\ref{fit_model}) in the illustrated time range. } \label{t2pp}
%\end{figure*}
%However, as will be discussed in the later
%context, the correlation functions do not show sizable contribution from these states except for glueball-like modes.
%On the other hand, our gluonic operators are built from different Wilson loops along with different smearing schemes,
%which can provide operators different enough from each other in the quenched approximation and therefore facilitate
%us to obtain optimal operators for the several lowest states. But in the presence of sea quark, some intrinsic dynamics
%of gluons and sea quarks may not differentiate these operators so much as that in the quenched approximation, such
%that the actual number of independent operators is much less. Anyway, this is still an open question to be
%investigate in the future.

Given the limited number of independent operators, our optimal operator $\Phi^{(R)}_i$ is actually not optimized as expected, namely, it does not only overlap to the $i$-th state but also to other states substantially. As seen in the effective mass plots, when $m_{1,{\rm eff}}^{(R)}(t)$ tends to reach a plateau as $t$ increases, $m_{2,{\rm eff}}^{(R)}(t)$ decreases gradually and finally merges into this plateau at large $t$ (within errors). Even though one can carry out the single exponential fit to the mass of the ground state in the plateau range
roughly beyond $t/a_t\approx 6 $ or 7, the bad signal-to-noise ratio in this time range results with large statistical errors. Since we focus on the ground states in the present study, in order to get more precise result of the masses
of the ground states, we adopt the following data-analysis strategy which also makes use of the measured data in the short time range. In each channel, we carry out a correlated fit to $\tilde{C}_1^{(R)}(t)$ and $\tilde{C}_2^{(R)}(t)$ simultaneously
through the following function forms,
\begin{eqnarray}
\label{fit_model}
\tilde{C}_1^{(R)}(t)&=&W_{11}^{(R)}e^{-m_1 t} + W_{12}^{(R)}e^{-m_2 t},\nonumber\\
\tilde{C}_2^{(R)}(t)&=&W_{21}^{(R)}e^{-m_1 t} + W_{22}^{(R)}e^{-m_2 t},
\end{eqnarray}
where the second mass term is introduced to take into account the contribution of the second state and higher states (of course, one can add more mass terms, but more parameters will ruin the data fitting due to the limited data points).
In the fitting procedure, the upper limit $t_{\rm max}$'s of the fit windows of $\tilde{C}_1^{(R)}(t)$ and $\tilde{C}_2^{(R)}(t)$ are chosen properly to include only the data points with good signal-to-noise ratios (The $t_{\rm max}$ of $\tilde{C}_2^{(R)}(t)$ are set to be from $7a_t$ to $9a_t$, while $t_{\rm max}$ of $\tilde{C}_1^{(R)}(t)$ can be larger than $10a_t$). 	Actually, the fit results are insensitive to $t_{\rm max}$'s in these ranges since they are almost determined by the data points in small $t$ range where relative errors are much smaller. For each channel, we keep $t_{\rm max}$'s fixed and vary $t_{\rm min}$ to check the stability and the quality of the fit.
The fit results for the scalar ($A_1^{++}$), the pseudoscalar ($A_1^{-+}$) and the tensor channels ( $E^{++}$ and $T_2^{++}$ ) at the two pion masses
are listed in Table~\ref{fit_result_high} and ~\ref{fit_result_low}. Except for $t_{\rm min}=1$ case in $T_2^{++}$ channel, all other
fits are acceptable with reasonable $\chi^2/d.o.f$. For all the four channels, the fitted parameters $m_1$ and $W_{11}$ are stable with
respect to the various $t_{\min}$, while $m_2$ decreases as $t_{\rm min}$ increasing gradually. This signals that our fitting model in Eq.~(\ref{fit_model}) is not so good that we should include more mass terms to account for higher states, which, however, affect the second state more than the first state. Since we are interested only in the first states, we do not take $m_2$ seriously and treat it as an object accommodating the effect of higher states.

In Fig.~\ref{a1pp},~\ref{a1mp},~\ref{epp} and~\ref{t2pp}, we also plot the shaded bands to illustrate the goodness of the fits. For each channel, after the correlated fit to the two correlations simultaneously, we get the six parameters $m_1$, $m_2$, $W_{11}^{(R)}$, $W_{12}^{(R)}$,$W_{21}^{(R)}$, and $W_{22}^{(R)}$ at different $t_{\rm min}$, which are listed in Table~\ref{fit_result_high} and~\ref{fit_result_low}. The red and blue bands are obtained through the function
	\begin{eqnarray}
	m_{i,{\rm eff}}^{(R)}(t)&=&\ln \frac{\tilde{C}_i^{(R)}(t)}{\tilde{C}_i^{(R)}(t+1)}\nonumber\\
	&=&
	\ln\frac{W_{i1}^{(R)}e^{-m_1 t} + W_{i2}^{(R)}e^{-m_2 t}}{W_{i1}^{(R)}e^{-m_1 (t+1)} + W_{i2}^{(R)}e^{-m_2 (t+1)}}.
	\end{eqnarray}
We calculate these values at each $t$ in the fit windows. The widths of the bands show the
errors estimated through the standard error propagation using the covariance error matrix of the parameters,
\[
\sigma_C^2 \approx \sum_{i,j=1}^6 \dfrac{\partial C}{\partial a_i}\sigma_{ij} \dfrac{\partial C}{\partial a_j},
\]
where $C$ denotes $m_{i,{\rm eff}}^{(R)}(t)$, $a_i$'s are the six parameters in Eq.~(\ref{fit_model}) and $\sigma_{ij}$'s are elements of covariance error matrix of the parameters, which are obtained directly from the fit. The extensions of the red and blue bands corresponds to the actual fit windows.

It is seen that the fit model describes the data of the ground state very well throughout the fit windows. For the second states, the fit model also fits the data more or less, especially in the small $t$ region. While in the large $t$ regions, the fitted results deviate somewhat from the data. This is understandable, since higher states, which do contribute, are missed in this model. This deviation actually contributes much to the $\chi^2$. It is expected that the fitted $m_2$ is generally (much) higher than the mass of the second state.
\begin{table*}[!htbp]
	\caption{Fitted results using fit model Eq.~(\ref{fit_model}) with different $t_{\rm min}$ at $m_\pi\sim 938~\mev$.  \label{fit_result_high} }
	\begin{center}
		\begin{tabular}{ccccccccc}
			\hline\hline
			$J^{PC}$ & $t_{\rm min}$ & $m_1a_t$  & $m_2a_t$ & $W_{11}$ & $W_{12}$ & $W_{21}$ & $W_{22}$ & $\chi^2/{\rm d.o.f}$\\
            \hline
			$A_1^{++}$&   1			 & 0.170(06) & 0.556(09)& 0.61(02) & 0.36(02) & 0.11(01) & 0.85(01) & 1.84				\\
			&    2         & 0.168(07) & 0.494(16)& 0.59(03) & 0.36(03) & 0.06(02) & 0.84(01) & 0.44               \\
			&     3         & 0.170(09) & 0.495(26)& 0.61(04) & 0.34(05) & 0.06(02) & 0.84(02) & 0.47               \\
			&     4         & 0.169(12) & 0.474(43)& 0.59(06) & 0.34(09) & 0.05(03) & 0.81(05) & 0.51               \\
			&     5         & 0.169(12) & 0.546(67)& 0.61(07) & 0.43(14) & 0.07(03) & 1.02(19) & 0.33               \\
%			&     5         & 0.161(15) & 0.464(83)& 0.55(09) & 0.44(08) & 0.04(03) & 0.79(24) & 0.52               \\
		    &&&&&&&&\\
			$A_1^{-+}$&	  1	& 0.307(09) & 0.720(15)& 0.66(02) & 0.33(03)  & 0.15(2)  &  0.79(2) &  2.51  \\
			&     2         & 0.316(13) & 0.665(28)& 0.67(05) & 0.28(05)  & 0.12(3)  &  0.79(2) &  1.21  \\
			&     3         & 0.306(19) & 0.633(38)& 0.62(08) & 0.33(10)  & 0.09(04) &  0.77(3) &  1.39  \\
			&     4         & 0.272(31) & 0.530(57)& 0.43(13) & 0.53(14)  & 0.02(5)  &  0.67(4) &  1.17  \\
%			&     5         & 0.261(40) & 0.494(76)& 0.37(18) & 0.56(20)  & -0.01(5) &  0.64(8) &  1.08  \\
%			&	  6			& 0.269(45) & 0.544(117)&0.43(21) & 0.56(37)  & 0.01(6)  &  0.78(28)&  1.19  \\
			&&&&&&&&\\
			$E^{++}$&	1	& 0.278(05) & 0.691(09)& 0.66(01) & 0.32(01)  & 0.19(1)  &  0.77(11) &  1.48  \\
			&     2         & 0.278(07) & 0.669(17)& 0.66(02) & 0.32(03)  & 0.18(2)  &  0.77(11) &  1.92  \\
			&     3         & 0.287(13) & 0.568(33)& 0.66(06) & 0.26(07)  & 0.12(4)  &  0.72(2) &  0.52  \\
			&     4         & 0.280(20) & 0.500(47)& 0.60(10) & 0.29(14)  & 0.05(6)  &  0.68(3) &  0.43  \\
			&     5        & 0.280(26) & 0.499(78)& 0.61(17) & 0.29(20)  & 0.06(8)  &  0.67(6) &  0.54  \\
%			&     5         & 0.296(30) & 0.512(123)& 0.74(23) & 0.05(37)  & 0.09(10) &  0.64(17) &  1.05  \\
			&&&&&&&&\\
			$T_2^{++}$&	 1  & 0.283(04) & 0.657(06)& 0.64(01) & 0.33(01)  & 0.15(09)  &  0.80(1) &  5.66  \\
			&     2          & 0.285(06) & 0.590(10)& 0.63(02) & 0.33(02)  & 0.10(1)  &  0.80(1) &  1.63  \\
			&     3         & 0.299(08) & 0.564(18)& 0.69(04) & 0.22(05)  & 0.09(2)  &  0.78(1) &  0.59  \\
			&     4         & 0.293(13) & 0.543(32)& 0.64(07) & 0.28(09)  & 0.06(3)  &  0.77(2) &  0.63  \\
			&     5         & 0.273(19) & 0.517(55)& 0.52(11) & 0.45(13)  & 0.04(5)  &  0.74(7) &  1.13  \\
%			&     5         & 0.260(29) & 0.491(74)& 0.45(16) & 0.53(19)  & 0.03(5)  &  0.64(10) &  1.20  \\
			\hline\hline
		\end{tabular}
	\end{center}
\end{table*}
\begin{table*}[!htbp]
	\caption{Fitted results using fit model Eq.~(\ref{fit_model}) with different $t_{\rm min}$ at $m_\pi\sim 650~\mev$ \label{fit_result_low} }
	\begin{center}
		\begin{tabular}{ccccccccc}
			\hline\hline
			$J^{PC}$ & $t_{\rm min}$ & $m_1a_t$  & $m_2a_t$ & $W_{11}$ & $W_{12}$ & $W_{21}$ & $W_{22}$ & $\chi^2/{\rm d.o.f}$\\
			\hline
			$A_1^{++}$&	  1			 &	0.176(09)& 0.548(14)& 0.62(03) & 0.36(03) & 0.09(02) & 0.85(02) & 1.71				\\
			&    2         & 0.163(14) & 0.481(23)& 0.55(05) & 0.42(06) & 0.04(2)  & 0.85(2)  & 0.98               \\
			&     3         & 0.173(17) & 0.515(37) & 0.59(07) & 0.38(09) & 0.06(3)  & 0.88(03) & 0.99               \\
			&     4         & 0.184(20) & 0.549(87)& 0.66(09) & 0.27(13) & 0.09(5)  & 0.92(15) & 1.12               \\
			&     5         & 0.150(25) & 0.533(99)& 0.49(11) & 0.69(22) & 0.06(4)  & 0.96(03) & 0.84               \\
			&&&&&&&&\\
			$A_1^{-+}$& 1   & 0.289(10) & 0.710(16)& 0.60(03) & 0.38(03)  & 0.12(2)  &  0.82(2) &  1.71  \\
			&     2         & 0.320(15) & 0.706(40)& 0.70(06) & 0.24(06)  & 0.15(4)  &  0.79(2) &  0.71  \\
			&     3         & 0.311(23) & 0.673(60)& 0.66(10) & 0.29(13)  & 0.13(5)  &  0.77(4) &  0.81  \\
			&     4         & 0.286(41) & 0.593(98)& 0.52(19) & 0.47(21)  & 0.07(8)  &  0.71(9) &  0.93  \\
%			&     5         & 0.188(73) & 0.448(61)& 0.15(14) & 0.84(12)  & -0.01(2) &  0.52(8) &  1.44  \\
			&&&&&&&&\\
			$E^{++}$& 1     & 0.268(07) & 0.646(10)& 0.62(02) & 0.36(02)  & 0.14(2)  &  0.81(1) &  1.42 \\
			&     2         & 0.267(10) & 0.601(20)& 0.61(04) & 0.35(04)  & 0.11(2)  &  0.80(1) &  0.81  \\
			&     3         & 0.255(13) & 0.578(28)& 0.55(05) & 0.42(06)  & 0.09(3)  &  0.79(1) &  0.71 \\
			&     4         & 0.254(20) & 0.563(49)& 0.54(09) & 0.42(11)  & 0.08(4)  &  0.77(6) &  0.85  \\
			&     5         & 0.272(36) & 0.453(86)& 0.62(24) & 0.17(31) &  0.00(11)&  0.66(2) &  0.46  \\
			&&&&&&&&\\
			$T_2^{++}$& 1   & 0.283(06) & 0.674(09)& 0.66(02) & 0.31(02)  & 0.18(1)  &  0.78(1) &  2.67  \\
			&    2          & 0.298(10) & 0.627(21)& 0.69(04) & 0.25(04)  & 0.15(3)  &  0.77(2) &  0.69  \\
			&     3         & 0.289(15) & 0.575(30)& 0.66(06) & 0.28(07)  & 0.11(4)  &  0.75(2) &  0.40  \\
			&     4         & 0.274(22) & 0.542(46)& 0.57(10) & 0.39(13)  & 0.08(5)  &  0.74(4) &  0.36  \\
			&     5         & 0.266(30) & 0.578(95)& 0.54(15) & 0.50(21)  & 0.09(6)  &  0.82(17) &  0.33  \\
			\hline\hline
		\end{tabular}
	\end{center}
\end{table*}

\begin{figure*}[!htbp]
	\centering
	\subfigure[$m_\pi\sim938~\mev$]{\label{average:900}\includegraphics[width=6.5cm]{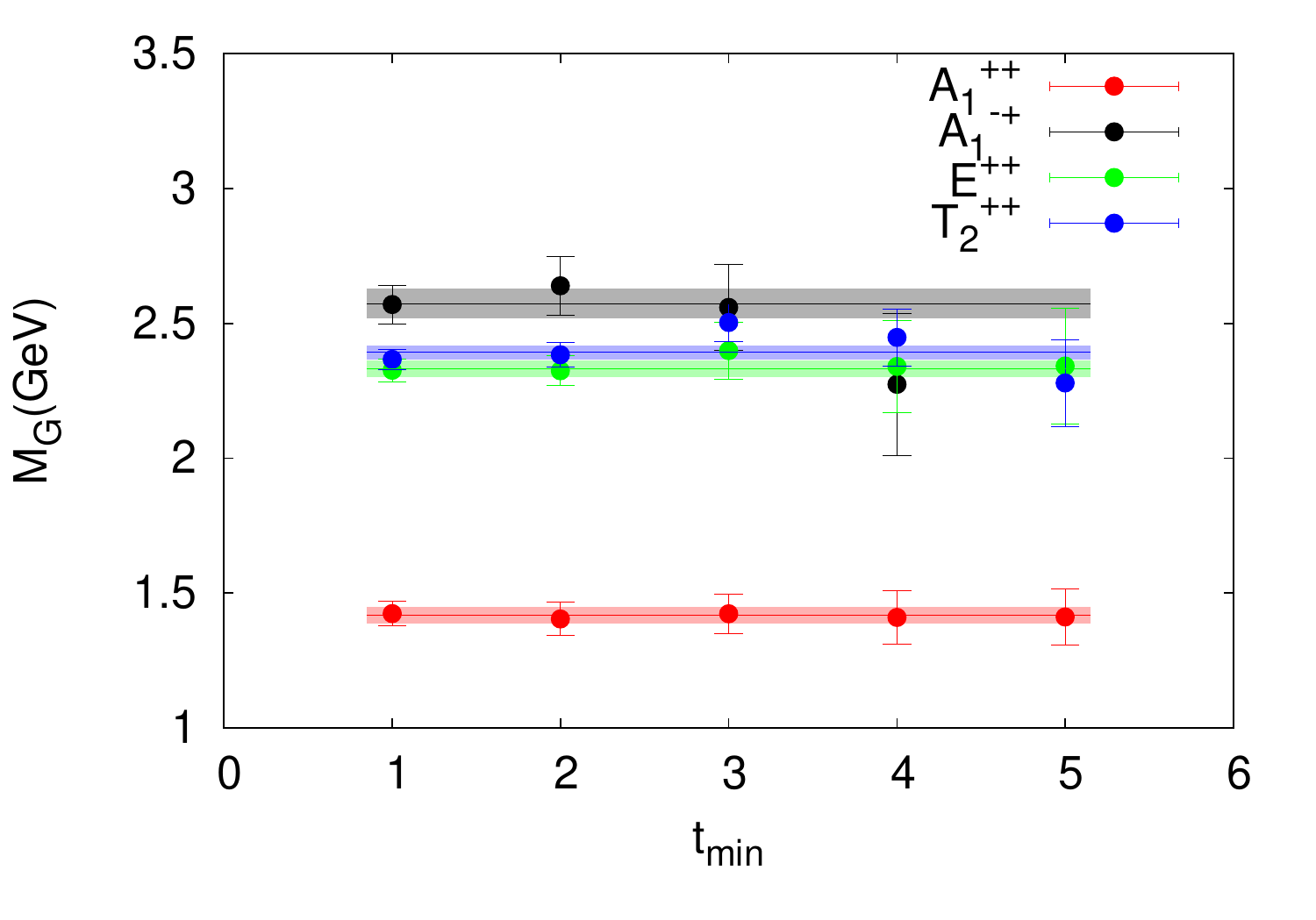}}\hfil
	\subfigure[$m_\pi\sim650~\mev$]{\label{average:600}\includegraphics[width=6.5cm]{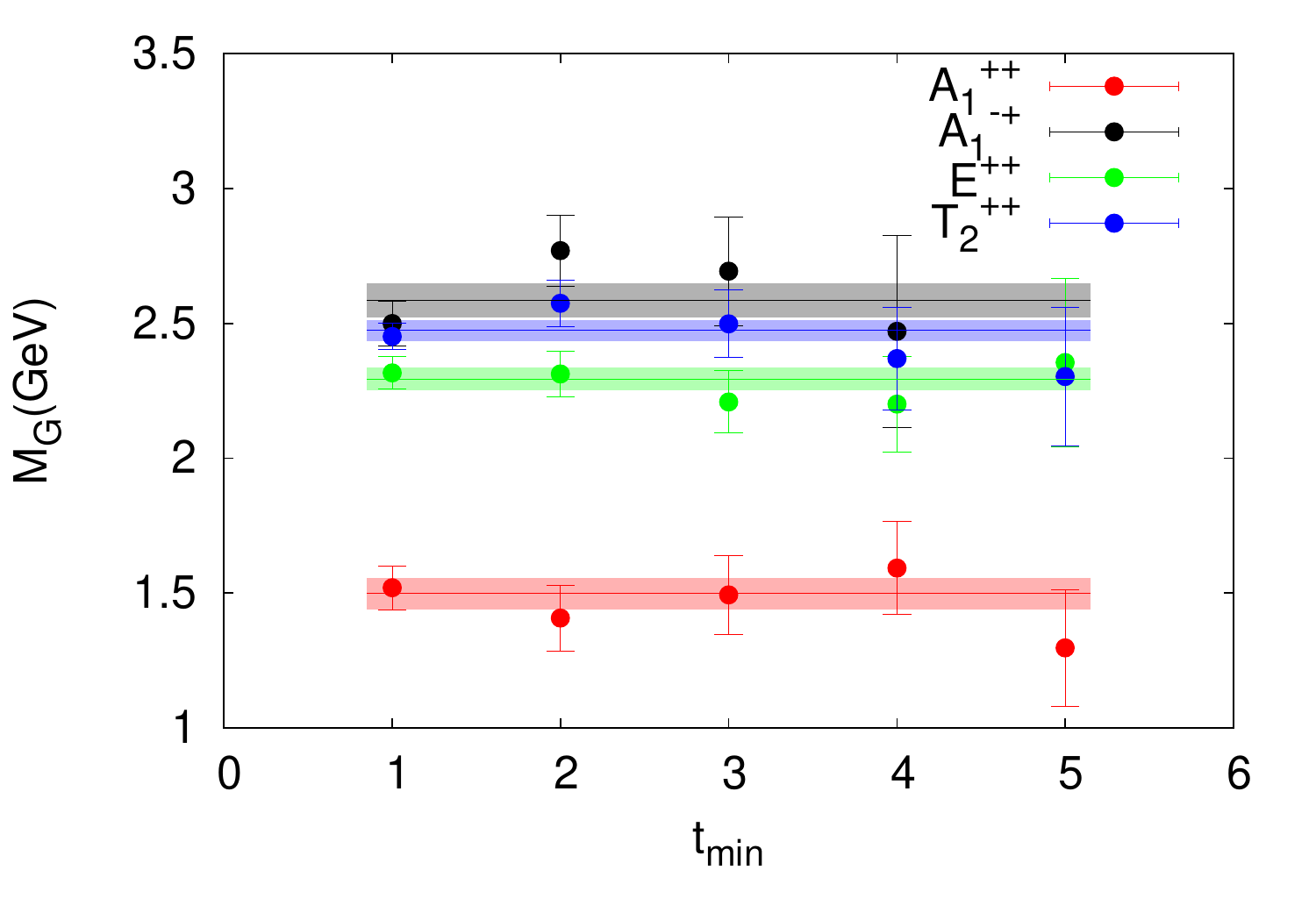}}
	\caption{Fitted $m_1$ for $A_1^{++},A_1^{-+},E^{++}$ and $T_2^{++}$ are plotted with respect to $t_{\rm min}$ (the left panel for $m_\pi\sim 938~\mev$ and the right panel for $m_\pi\sim650~\mev$). The values are expressed in the physical units inverted by the lattice spacing listed in Table~\ref{conf details}. The color bands illustrate the averaged values weighted by the inversed squared errors at each $t_{\rm min}$.  } \label{average}
\end{figure*}

As shown in Table~\ref{fit_result_high} and ~\ref{fit_result_low}, most of the fits using different $t_{\rm{min}}$ are statistically acceptable and
the masses of the first states are relatively stable. Therefore, for the final result of $m_1$ in each channel, we take tentatively the average value of $m_1$'s at different $t_{\rm{min}}$ weighted by their inversed squared errors. The statistical errors are accordingly derived. This averaging is illustrated in Fig.~\ref{average}, where data points are the fitted result of $m_1$ at different $t_{\rm{min}}$ and the shaded bands are the averaged values with averaged errors. The results are also listed in Table~\ref{mass_values}. At the heavy pion mass $m_\pi\sim 938$ MeV, $m_1(E^{++})$
is very close to $m_1(T_2^{++})$, as expected by the rotational symmetry restoration in the continuum limit where they correspond to the mass of the
same $2^{++}$ tensor state. However for the lighter $m_\pi\sim 650$ MeV, the two masses deviate from each other by 200 MeV. Since the lattice
spacings at the two pion massed are very close, the extent of the rotational symmetry breaking should be similar. We tentatively attribute this large
deviation to the relatively small statistics at $m_\pi\sim 650$ MeV, which is roughly one-half as large as that at $m_\pi\sim 938$ MeV (see Table ~\ref{conf details}). From Table~\ref{meson_spec} and Table~\ref{mass_values} one can see that the masses of ground state scalar meson and our scalar glueball are very close to each other, this may indicate there are mixing between $q\bar{q}$ and the scalar glueball, which needs further investigation.

\begin{table*}[htbp]
\caption{\label{mass_values} Final results for the masses of the lowest state we obtain in the $A_1^{++},A_1^{-+},E^{++}$ and $T_2^{++}$ channels.
These are the averaged values weighted by the inversed squared errors at each $t_{\rm min}$. }
\begin{center}
\begin{tabular}{ccccc}
\hline
\hline
$m_\pi$(MeV)   &   $m_1(A_1^{++})$(MeV)   &      $m_1(E^{++})$(MeV)   &   $m_1(T_2^{++}) $(MeV)   &   $m_1(A_1^{-+})$(MeV)   \\\hline
$938 $    &   1417(30)               &    2332(31)               &   2392(26)                &   2573(55) \\
$ 650 $    &   1498(58)               &    2294(43)               &   2474(39)                &   2585(65) \\
\hline
\hline
\end{tabular}
\end{center}
\end{table*}

\subsection{Interpretation of the ground states}

Generally speaking, the two-point function of an interpolating operator $\mathcal{O}(t)$ with definite quantum numbers is usually parameterized as
\begin{equation}\label{parameterization}
 C(t)=\langle 0|\mathcal{O}(t)\mathcal{O}^\dagger(0)|0\rangle=\sum\limits_n\langle 0|\mathcal{O}|n \rangle\langle n|\mathcal{O}^\dagger|0\rangle e^{-m_n t},
\end{equation}
where $\{|n\rangle, n=1,2,\ldots\} $ are eigenstates of Hamiltonian with eigenvalue $m_n$, which make up an orthogonal, normalized, and complete state set with
\begin{equation}
\sum\limits_n |n\rangle\langle n|=1,~~~~\langle n|n'\rangle=\delta_{nn'}.
\end{equation}
For QCD on a Euclidean spacetime lattice, $m_n$ take discretized values and
the connection of these discretized energy levels to the relevant $S$-matrix parameters
should be established through other theoretical formalisms, such as L\"{u}scher's.
Here we would only focus on the physical meaning of the
fitted masses of the lowest states.

We take the scalar channel for instance. A hadron system of the bare states with the scalar quantum number $J^{PC}=0^{++}$ can be a bare scalar glueball $|G_{0^{++}}\rangle$, a bare $q\bar{q}$ scalar meson $|f_0\rangle$, or even $\pi\pi$ scattering states $|\pi\pi\rangle$.
We simplify the matter further by assuming that the two adjacent states mix most,
then we can only consider a two-state system composed of the ground state scalar glueball $|G\rangle$
and its adjacent state, which could be of nature $|\pi\pi\rangle$ or $|f_0\rangle$.
This then yields the fitting model in Eq.~(\ref{fit_model}) that we introduced previously.

We compare the results in the present study with the previous quenched and
unquenched results in Table~\ref{compare}. The tensor glueball masses are obtained by averaging the corresponding
$E^{++}$ and $T_2^{++}$ values.
Despite the fact that glueball correlation functions in the unquenched QCD acquire more complicated spectrum decomposition than the quenched case, the mass of the bare glueball states $|G\rangle$ can still be obtained
by assuming the corresponding operators $\mathcal{O}$ couple weakly to other states.
Therefore, it is naturally understood that the glueball spectrum in our full-QCD lattice studies is similar to that in the quenched approximation. The difference is still visible, however, and it is most evident in the scalar channel where one would expect that this weak coupling assumption is not valid anymore.
\begin{table*}[tbp]
\caption{\label{compare} We compare our results with previous results both from the quenched lattice QCD studies~\cite{Morningstar:1999rf, Chen:2005mg}
and the full-QCD study~\cite{Gregory:2012hu}. We average the masses of $E^{++}$ and $T_2^{++}$ states to obtain the estimate of the $2^{++}$ glueball mass.}
\begin{center}
\begin{tabular}{lclll}
\hline\hline
                              &      $m_\pi$ (MeV)    & $m_{0^{++}}$ (MeV)     &   $m_{2^{++}}$ (MeV)     &   $m_{0^{-+}}$ (MeV)      \\
\hline

$N_f=2$                       &  $938$  &    1417(30)    &    2363(39)             &   2573(55)               \\
                              &  $650$  &    1498(58)    &    2384(67)             &   2585(65)               \\
%%&&&&\\
$N_f=2+1$~\cite{Gregory:2012hu}&  $360$ &    1795(60)    &    2620(50)             &    ---                    \\
&&&&\\
quenched~\cite{Morningstar:1999rf} &        ---          &    1710(50)(80)&    2390(30)(120)        &    2560(35)(120)         \\
quenched~\cite{Chen:2005mg}        &        ---          &    1730(50)(80)&    2400(25)(120)        &    2590(40)(130)         \\

\hline\hline
\end{tabular}
\end{center}
\end{table*}

\section{Further study on the pseudoscalar channel}
\label{sec:pseudoscalar}

As presented in the last section, in the $A_1^{-+}$ channel, we obtain the
mass of the ground state to be $m_{A_1^{-+}}\sim 2.6$ GeV at the two pion masses,
which is compatible with the pure gauge glueball mass.
Theoretically, in the presence of sea quarks, the flavor singlet $q\bar{q}$ pseudoscalar meson
is expected to exist, but we do not observe this state from  the correlation function of the glueball
operator $\Phi^{(PS)}$.

In order to check the existence of the flavor singlet pseudoscalar meson in the spectrum, we would like to
study the correlation function of topological charge density operator $q(x)$. 
This is motivated by the partially conserved axial current (PCAC),
\begin{equation}
\partial_\mu
J_5^{\mu}(x)=2mP(x)-\frac{N_f g^2}{16\pi^2}\epsilon_{\mu\nu\rho\sigma} Tr F^{\mu\nu}F^{\rho\sigma},
\end{equation}
where $g$ is the strong coupling constant, $P(x)=\bar{\psi}(x)\gamma_5 \psi(x)$ is the pseudoscalar density,
and the anomalous gluonic operator $\epsilon_{\mu\nu\rho\sigma}F^{\mu\nu}F^{\rho\sigma}$ is the so-called
topological charge density (up to a constant factor), which is usually denoted by $q(x)$. 
Thus $q(x)$ may have substantial overlap with the flavor
singlet pseudoscalar meson (denoted by $\eta^\prime$).

The correlation function of $q(x)$ is expressed as
\begin{equation}
C_q(x-y)=\langle q(x)q(y)\rangle,
\end{equation}
from which one can get the topological susceptibility 
\begin{equation}
	\chi_t=\frac{1}{V_4}\int d^4x d^4 y C_q(x-y),
\end{equation}
where $V_4$ is the four-dimensional volume
of the Euclidean spacetime. It is known that $\chi_t$ is positive and takes a value
$\sim(180~MeV)^4$. On the other hand, $q(x)$ is a pseudoscalar operator and requires $C_q(x-y)<0$
for $r=||x-y||>0$. So $C_q(x-y)$ can be intuitively expressed as
\begin{equation}
C_q(x-y)=A\delta^4(x-y)+\bar{C}_q(x-y),
\end{equation}
where $\bar{C}_q(x-y)$ is negative for $r>0$. On the Euclidean spacetime lattice with a finite lattice spacing, the delta function will show up a positive kernel with a width of a few lattice spacings, and $C_q(x-y)$ has a negative tail contributed from $\bar{C}_q(x-y)$.
It is expected that $\bar{C}_q(x-y)$ would be dominated by the contribution of the lowest
pseudoscalar meson in the large $r$ range and can be parameterized as~\cite{PhysRevD.52.295}
\begin{equation}\label{correform}
\bar{C}_q(r)=N\frac{m_{\rm PS}}{4\pi^2r}K_1(m_{\rm PS}r),
\end{equation}
where $N$ is an irrelevant normalization factor, $m_{\rm PS}$ is the mass of the lowest pseudoscalar, and $K(z)$ is the modified Bessel function of second kind, whose asymptotic form at large $|z|$ is
\begin{equation}
K_1(z)\sim\sqrt{\frac{\pi}{2z}}e^{-z}(1+\frac{3}{8z}),\quad |arg~z|<\frac{3}{2}\pi.
\end{equation}
Therefore, one can obtain $m_{\rm PS}$ by fitting the negative tail of $C_q(x-y)$ in the large $r$ range using the above functional form. 

This has been actually done by several lattice studies in both the quenched approximation~\cite{Chowdhury:2014mra} and full QCD calculations~\cite{Fukaya:2015ara}. In the quenched approximation,
the extracted $m_{\rm PS}=2563(34)$ MeV is in good agreement with the pseudoscalar glueball mass $m_{\rm PS}=2560(35)$ MeV. This is as it should be, since
the hadronic excitations of a pure gauge theory are only glueballs. In the full-QCD study with $N_f=2+1$ and pion masses close to the physical $m_\pi$, $m_{\rm PS}$ is obtained to be $1013(117)$ MeV, which is consistent with the mass of the physical $\eta'$.
\begin{figure*}[thbp]
	\centering
	\subfigure[$m_\pi\sim938~\mev$]{\includegraphics[width=6.5cm]{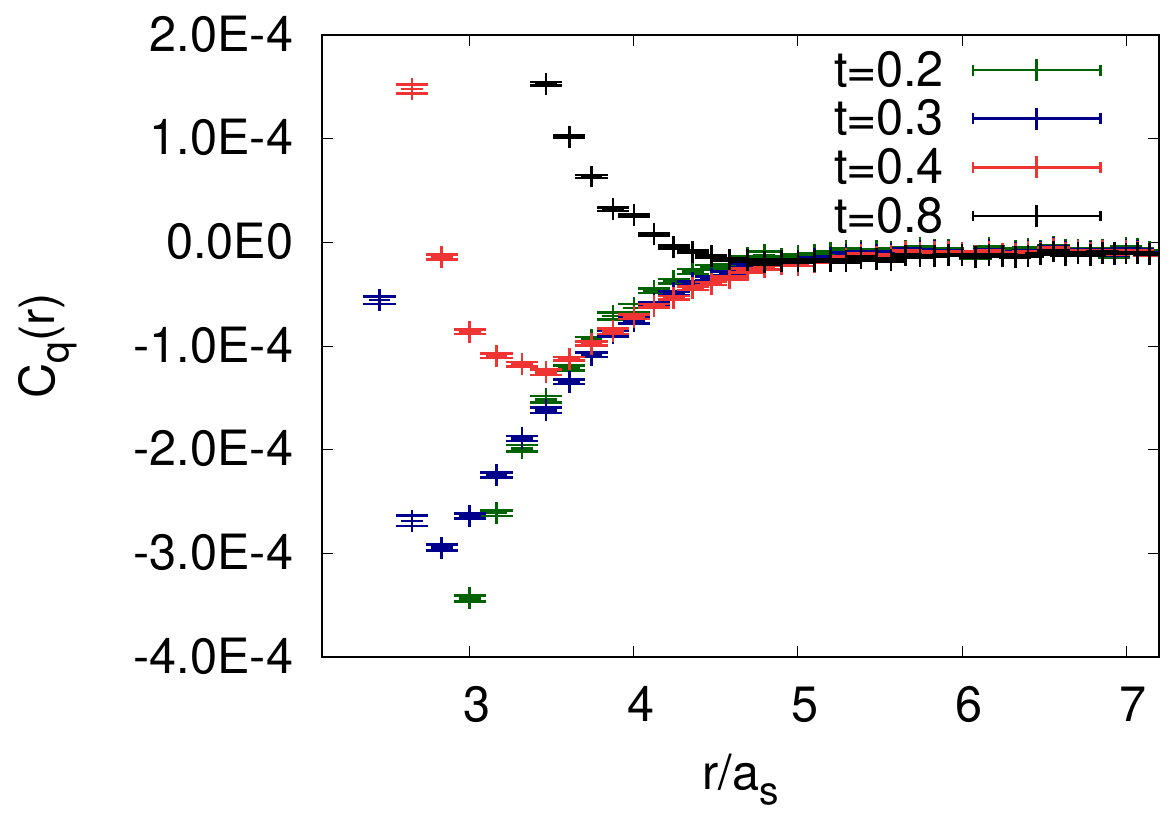}}\hfil
	\subfigure[$m_\pi\sim650~\mev$]{\includegraphics[width=6.5cm]{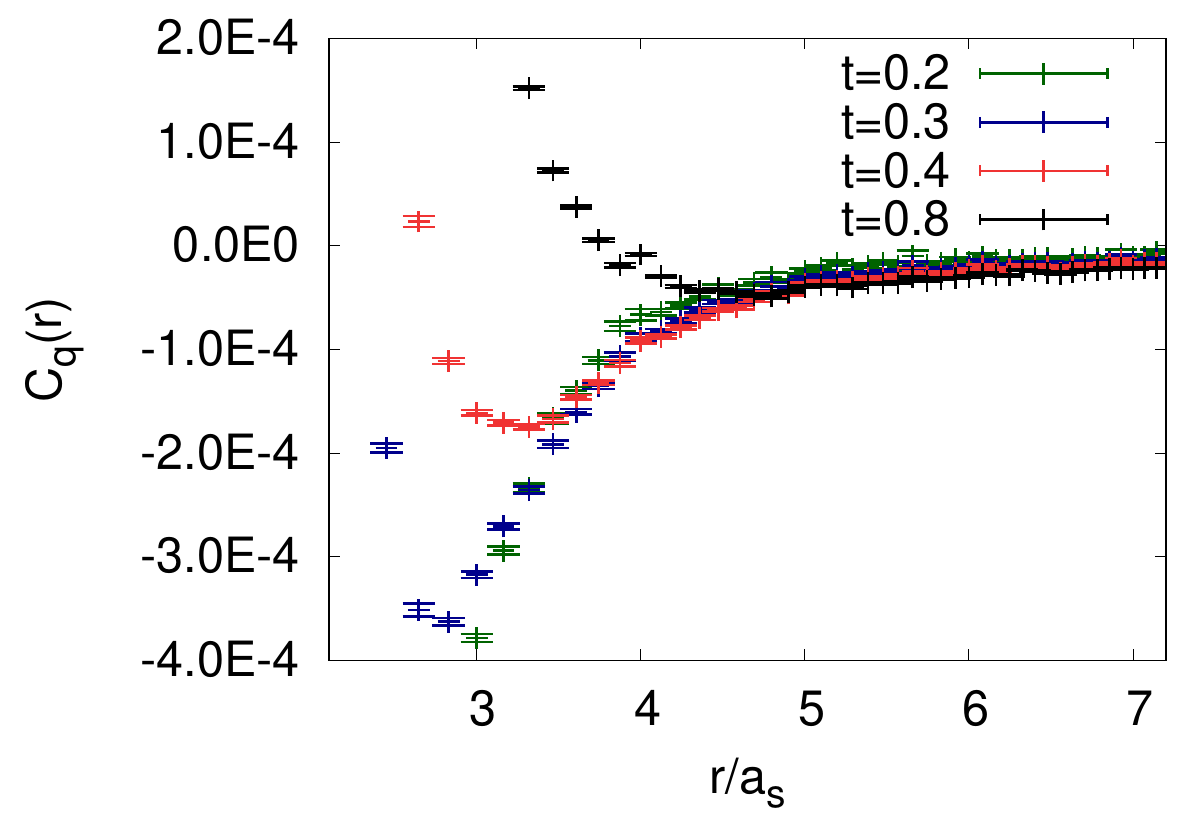}}
	\caption{\label{twopf_eta}The correlation function $C_q(r)$ of topological charge
		density in terms of the four dimensional Euclidean distance (the left panel for $m_\pi\sim 938~\mev$ and the right panel for $m_\pi\sim650~\mev$. Different curves correspond to $C_q(r)$ at different Wilson flow time $t=0.2,0.3,0.4$ and $0.8$.}
\end{figure*}
In this work, we adopt a similar strategy to that in~\cite{Fukaya:2015ara}. The topological charge density $q(x)$ is defined by the spatial and temporal
Wilson loops (plaquettes) as conventionally done. We use the  Wilson gradient flow method as a smearing scheme to optimize the behavior of topological charge density correlator~\cite{Moran:2008qd,Fukaya:2015ara}. The Wilson flow provides a reference energy scale
$\frac{1}{\sqrt{8t}}$~\cite{Luscher:2010iy}. In practice, we use the code published by the BMW collaboration~\cite{Borsanyi:2012zs}
to evaluate the topological charge density. Fig.~\ref{twopf_eta} shows $C_q(r)$ for
$m_\pi\sim938~\mev$ and $m_\pi\sim650~\mev$ at flow times $t=0.2,0.3,0.4,0.8$ respectively. On our lattices, these $t$ values correspond to $\sqrt{8t}\sim0.15,0.18,0.21$ and $0.30~\fm$. As shown in the figures, at large flow time, $C_q(r)$ is mostly positive, which implies that the gauge fields are over smeared. 

\begin{figure*}[tbp]
	\centering
	\subfigure[$m_\pi\sim938~\mev$]{\includegraphics[width=6.5cm]{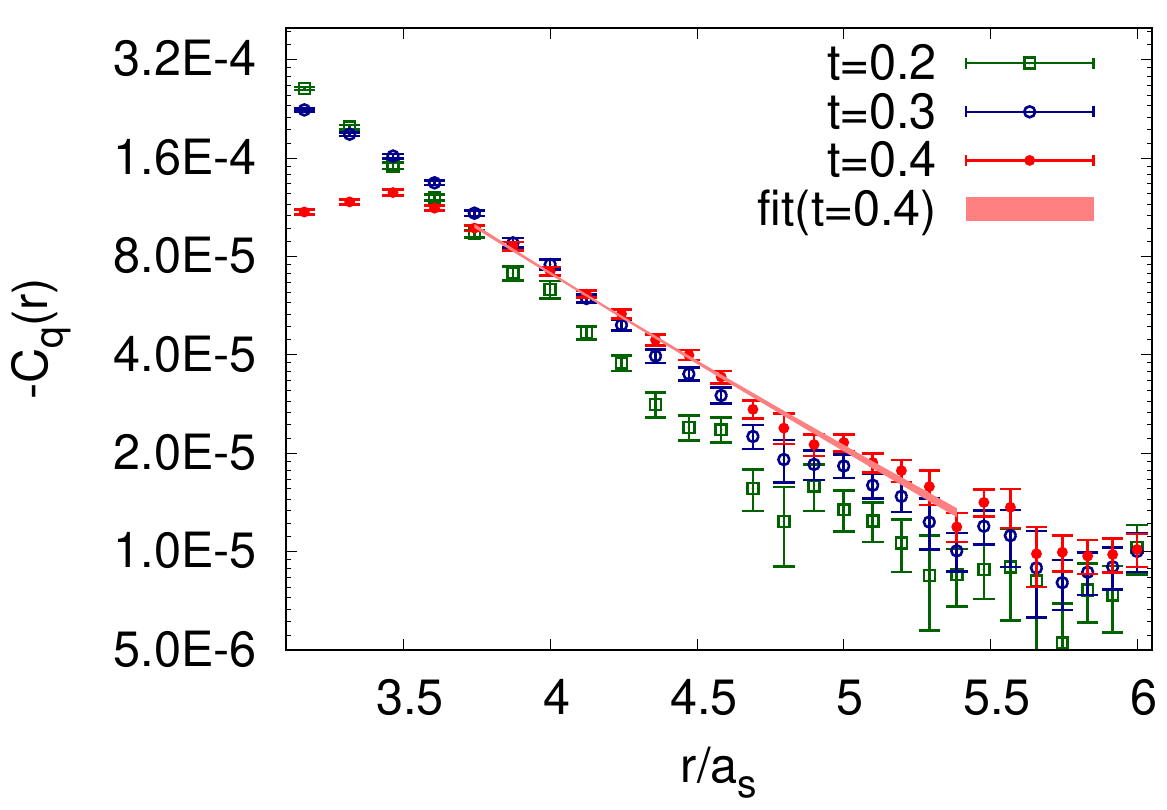}}\hfil
	\subfigure[$m_\pi\sim650~\mev$]{\includegraphics[width=6.5cm]{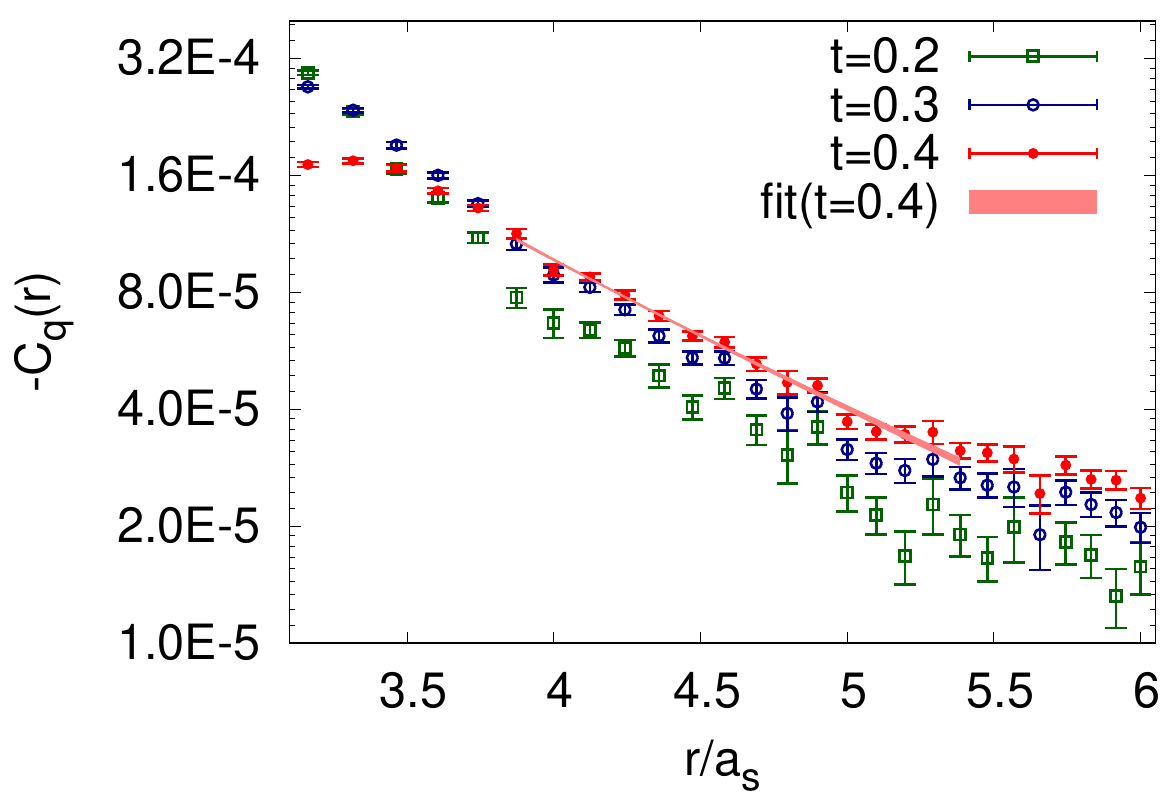}}
	\caption{\label{effecmass_eta}-$C_q(r)$ at different flow times $t=0.2,0.3,0.4$ are plotted in log scale
		for comparison in the large $r$ range, for $m_\pi\sim938$ MeV (left) and $m_\pi\sim650~\mev$ (right). 
		In each panel, the red band illustrates the fit to $C_q(r)$ at the flow time $t=0.4$ in the $r$
		interval $r/a_s\in[3.8,5.4]$.}
\end{figure*}

In order to compare the large $r$ behaviors of $C_q(r)$ at different flow times, we plot them in Fig.~\ref{effecmass_eta} in logarithmic scale, where one can see that their behaviors are similar in the 
large $r$ region, but the $C_q(r)$ at $t=0.4$ looks the smoothest and has the smaller errors.
Therefore, we fit the $C_q(r)$ at $t=0.4$ directly through the function form of Eq.~\ref{correform} to extract
the parameter $m_{PS}$. In determining the fit range, we take the following two factors into consideration.
First, the spatial extension of our lattices is $L_s=12a_s$.
In order to avoid large finite volume effects, the upper limit of the fit range should be smaller than $6a_s$,
due to the periodic spatial boundary condition.
Secondly, as shown in Fig.~\ref{twopf_eta}, the negative tail of $C_q(r)$ starts beyond $r\sim3a_s$,
which requires the lower limit of the fit range to be larger than $3a_s$. In the practical fitting procedure
of $C_q(r)$ at $t=0.4$, we choose the fit range to be $r/a_s\in[3.8,5.4]$.

We carry out a correlated minimal-$\chi^2$ fit to $C_q(r)$ at $t=0.4$ in the $r$ interval described above.
Table~\ref{effc_eta_tab} lists the fit ranges, the fitted results of $m_{PS}$ and the $\chi^2/dof$'s at the
two pion masses. In order to illustrate the fit quality, we also plot $C_q(r)$ in Fig.~\ref{effecmass_eta}
in red bands using the function form in Eq.~\ref{correform} with the fitted parameters. The $m_{PS}$'s we get
are around 1 GeV and show explicit dependence on the pion mass. However, they are much smaller than the values
around 2.6 GeV from the correlation functions of the pseudoscalar glueball operator $\Phi^{PS}$. Thus the light
pseudoscalar state observed in $C_q(r)$ can be naturally assigned to be the flavor singlet $q\bar{q}$ state
$\eta^\prime$. Theoretically, the mass of $\eta'$ is acquired through the interaction of sea quark loops according to the Witten-Veneziano mechanism~\cite{Witten:1979vv,VENEZIANO1979213}. In this mechanism, the propagator of 
$\eta^\prime$ can be expressed as
\begin{eqnarray}
\frac{1}{q^2-m_{\eta'}^2} &=& \frac{1}{q^2-m_\pi^2}\left(1+m_0^2\frac{1}{q^2-m_\pi^2} \right. \nonumber\\
& &\left. +m_0^2\frac{1}{q^2-m_\pi^2}m_0^2\frac{1}{q^2-m_\pi^2}+\ldots\right),
\end{eqnarray}
where the parameter $m_0^2$ is introduced to describe the gluonic coupling, such that
\begin{equation}
m_{\eta'}^2\approx m_\pi^2+m_0^2.
\end{equation}
On the other hand, $m_0^2$ is related to the topological susceptibility $\chi_t$ through
\begin{equation}\label{WV}
m_0^2=\frac{4N_f}{f_\pi^2}\chi_t,
\end{equation}
where $f_\pi$ is the decay constant of $\pi$. For our case of $N_f=2$, if we take the values $\chi_t=(180\,{\rm MeV})^4$, $f_\pi\sim 150$ MeV for $m_\pi\sim 650$ MeV and $f_\pi\sim 200$ MeV for $m_\pi\sim938$ MeV, $m_0^2$ is estimated to be approximately $(610\,{\rm MeV})^2$ and $(460\,{\rm MeV})^2$, respectively. Thus the $\eta'$ mass can be derived as $m_{\eta'}\sim 890$ MeV for $m_\pi\sim650$ MeV, and $m_{\eta'}\sim 1045$ MeV for $m_\pi\sim938$ MeV. These values are not far from the $m_{\rm PS}$'s we obtained.

Because these are very preliminary calculations and the systematic errors are not well under control, 
we do not want to overclaim the values of $m_{PS}$ we obtain. What we would like to emphasize is that there does exist in the spectrum a flavor singlet $q\bar{q}$ pseudoscalar meson corresponding to the $\eta^\prime$ meson
in the real world, which can be accessed by the topological charge density operator.
\begin{table}[tbp]
	\caption{\label{effc_eta_tab} The fitting details for $\eta^\prime$ meson mass from topological
		charge density correlator at $m_\pi\sim938~\mev$ and $m_\pi\sim650~\mev$ respectively.}
	\begin{center}
		\begin{tabular}{c|c|c|c|c}
			\hline \hline
			$m_\pi$             &     fit range($a_s$)   &   $m_{\eta'}a_s$          &   $m_{\eta^\prime}(\mev)$  &   $\chi^2/dof$    \\\hline
			$938~\mev$        &     3.74-5.92          &   $0.856(21)$     &   $1481(36)$             &   $1.01$  \\
			$650~\mev$        &     3.87-5.48            &   $0.514(22)$     &   $890(38)$              &   $1.43$\\\hline \hline
		\end{tabular}
	\end{center}
\end{table}

Now that the $\eta'$ state exists in the spectrum, there comes a question of why it is missing in the correlation function of the conventional gluonic operator for the pseudoscalar glueball (denoted as $\Phi^{\rm PS}$).
In order to clarify this, we check the continuum form of $\Phi^{({\rm PS})}$ involved in this work. Actually, in the construction of the
gluonic pseudoscalar operators, only the spatially solid (instead of planar) Wilson loops (the last four prototypes in Fig.~\ref{loops}) are used,
\begin{eqnarray}
\phi_{\alpha}^{A_1^{-+}}(\mathbf{x},t) &=&\sum_{R\in O} c_{R}^{A_1^{-+}}ReTr\left[R\circ
W_{\alpha}(\mathbf{x},t)\right.\nonumber \\
& &-\left.\mathcal{P}R\circ W_{\alpha}(\mathbf{x},t)\mathcal{P}^{-1}\right],
\end{eqnarray}
where $R$ stands for each rotation operation in $O$ group, $c_{R}^{A_1}$ is the combinational
coefficients corresponding to the $A_1$ irreducible representation, $W_\alpha$ is any of the four
prototypes made up of a specifically smeared gauge links. According to the non-Abelian Stokes
theorem~\cite{Simonov:1988xs}, a rectangle Wilson loop $P_{\mu\nu}^{a\times b}(x)$ of size $a\times b$,
with $a,b$ small, can be expanded as
\begin{widetext}
\begin{eqnarray}
\label{expansion} P_{\mu\nu}^{a\times b}(x) &=& {\bf 1} +
ab(F_{\mu\nu}(x)+\frac{1}{2}(aD_{\mu}+bD_\nu)F_{\mu\nu}(x)\nonumber\\
&+&\frac{1}{12}(2a^2D_\mu ^2+3abD_\mu D_\nu+2b^2
D_{\nu}^2)F_{\mu\nu}(x)\nonumber\\
&+&\frac{1}{24}(a^3D_{\mu}^3 + 2a^2 b D_{\mu}^2 D_{\nu} + 2 ab^2
D_{\mu}D_{\nu}^2 + b^3 D_{\nu}^3)F_{\mu\nu}(x))\nonumber\\
&+&(ab)^2(\frac{1}{2}F_{\mu\nu}^2(x) +
\frac{1}{2}F_{\mu\nu}(x)(aD_{\mu}+bD_{\nu})F_{\mu\nu}(x)\nonumber\\
&+& \frac{1}{24}F_{\mu\nu}(x)(a^2 D_{\mu}^2+b^2
D_{\nu}^2)F_{\mu\nu}(x))+\frac{1}{6}(ab)^3 F_{\mu\nu}^3 + O((ab)^4).
\end{eqnarray}
\end{widetext}
%\begin{eqnarray}
%\label{expansion} P_{\mu\nu}^{a\times b}&(x)& = {\bf 1} +
%ab(F_{\mu\nu}(x)+\frac{1}{2}(aD_{\mu}+bD_\nu)F_{\mu\nu}(x)\nonumber\\
%&+&\frac{1}{12}(2a^2D_\mu ^2+3abD_\mu D_\nu+2b^2
%D_{\nu}^2)F_{\mu\nu}(x)\nonumber\\
%&+&\frac{1}{24}(a^3D_{\mu}^3 + 2a^2 b D_{\mu}^2 D_{\nu} + 2 ab^2
%D_{\mu}D_{\nu}^2 + b^3 D_{\nu}^3)F_{\mu\nu}(x))\nonumber\\
%&+&(ab)^2(\frac{1}{2}F_{\mu\nu}^2(x) +
%\frac{1}{2}F_{\mu\nu}(x)(aD_{\mu}+bD_{\nu})F_{\mu\nu}(x)\nonumber\\
%&+& \frac{1}{24}F_{\mu\nu}(x)(a^2 D_{\mu}^2+b^2
%D_{\nu}^2)F_{\mu\nu}(x))\nonumber\\
%&+&\frac{1}{6}(ab)^3 F_{\mu\nu}^3 + O((ab)^4),
%\end{eqnarray}
where $F_{\mu\nu}$ is the strength tensor of the gauge field.
For simplicity, the factor $ig$ is absorbed into the quantity $F_{\mu\nu}$. The small $ab$
expansion of $ P_{\pm\mu\pm\nu}(x)$ is similar to Eq.~\ref{expansion} by replacing $a$ and $b$
with $\pm a$ and $\pm b$, respectively. Since the last four prototypes can be expressed as products of
two rectangle Wilson loops, using the above relation one can obtain the leading term of the
pseudoscalar operator,
\begin{equation}
\phi_\alpha^{A_1^{-+}}(\mathbf{x},t)\propto \epsilon_{ijk}TrB_i(\mathbf{x},t)D_j
B_k(\mathbf{x},t)+O(a_s^2),
\end{equation}
which is obviously different from the anomalous part of the PCAC relation,
$\epsilon_{\mu\nu\rho\sigma} F^{\mu\nu}(x)F^{\rho\sigma}(x)\propto
\mathbf{E}(x)\cdot\mathbf{B}(x)$. Actually, the operator $\Phi^{({\rm PS})}$ is a linear
combination of these kinds of operators defined through differently smeared gauge fields.
This may imply that the two operators couple differently to
specific states. Along with the observation
in the calculation of glueball spectrum, this proves to some extent that our operator for the pseudoscalar glueball couples very weakly to the $q\bar{q}$ meson state and almost exclusively to the glueball states.

\begin{table*}[tbp]
	\centering
	\caption{\label{pseudo} The table collects the masses of flavor singlet pseudoscalar mesons from the quenched and unquenched lattice QCD studies. $P(x)$, $q(x)$ and $\Phi^{({\rm PS})}$ stand for the quark bilinear pseuscalar operator, the topological charge density, and the pseudoscalar glueball operator, respectively.}
	\begin{center}
		\begin{tabular}{l|ccc}
			\hline
			\hline
			&		$P(x)$	&  $q(x)$	 &  $\Phi^{({\rm PS})} $ \\\hline
			$N_f=0$	&	 	------	&{\bf 2563(34)} MeV~\cite{Chowdhury:2014mra}	& {\bf 2590(140)} MeV~\cite{Chen:2005mg}\\
			$N_f=2$	&	768(24) MeV~\cite{Helmes:2017ccf}&	890(38) MeV (this work) 	&	{\bf 2585(65)} MeV  (this work)	\\
			$N_f=2+1$&	947(142) MeV~\cite{Christ:2010dd}	& 1019(119) MeV~\cite{Fukaya:2015ara}	&	------				\\
			$N_f=2+1+1$&	{1006(65)MeV}	~\cite{Michael:2013gka}&	------	   &	------		\\
			\hline\hline
		\end{tabular}
	\end{center}
\end{table*}

We collect the existing lattice results of the masses of flavor singlet pseudoscalar mesons in Table~\ref{pseudo} for an overview. In the quenched approximation ($N_f=0$), the authors
of Ref.~\cite{Chowdhury:2014mra} use $q(x)$ as pseudoscalar operators and derive the ground state mass $m_{\rm PS}=2.563(34)$ GeV, which is almost the same as the mass of the pure gauge pseudoscalar glueball $m_{PS}=2.560(140)$ GeV~\cite{Morningstar:1999rf} and $2.590(140)$ GeV~\cite{Chen:2005mg}.  This is exactly what it should be, since there are only pseudoscalar glueball propogating along time if no valence quarks are involved.
When dynamical quarks are included in the lattice simulation, the situation is totally different. There have been several works using $P(x)$ to calculate the
$\eta'$ mass in the lattice simulation with dynamical quarks, and have given the results $m_{\eta'}=768(24)$ MeV ($N_f=2$)~\cite{Helmes:2017ccf}, $m_{\eta'}=947(142)$ MeV ($N_f=2+1$)~\cite{Christ:2010dd} and $m_{\eta'}=1006(65)$ MeV ($N_f=2+1+1$)~\cite{Michael:2013gka}, which almost reproduce the experimental result $m_{\eta'}=958$ MeV. When the $q(x)$ operator is applied, $N_f=2+1$ lattice simulation gives the result $m_{\eta'}=1019(119)$ MeV at the physical
pion mass~\cite{Fukaya:2015ara}, which is consistent with the result through the $P(x)$ operator. We also calculate the ground state mass using the $q(x)$ operator
on our $N_f=2$ gauge configurations and obtain the result $m_{\rm PS}=890(38)$ MeV at $m_\pi=650$ MeV, which is compatible with the $m_{\eta'}=768(24)$ MeV above (note that our $m_\pi$ is higher than that in Ref.~\cite{Helmes:2017ccf}). The similar result for $m_{\eta'}$ from the operators $P(x)$ and $q(x)$ can be understood as follows. Due to the $U_A(1)$ anomaly, $q(x)$ is now related to $P(x)$ through the PCAC relation. The relation implies that $q(x)$ can couple substantially to the flavor singlet $\eta'$ meson. In contrast, the glueball operator $\Phi^{({\rm PS})}$ couples predominantly to the pseudoscalar glueball
state either in the quenched approximation or in the presence of sea quarks.

\section{Summary and conclusions}
\label{sec:summary}

The spectrum of the lowest-lying glueballs is investigated in lattice QCD with two flavors of degenerate Wilson clover-improved quarks. We generate ensembles of gauge configurations on anisotropic lattices at two pion masses, $m_\pi\sim650$ MeV and
$m_\pi\sim938$ MeV. Focus has been put on the ground states of the scalar, pseudoscalar and tensor glueballs, which are measured using gluonic operators constructed from different prototypes of Wilson loops. The variational method is applied to obtain the optimal operators which couple dominantly to the ground state glueballs.

In the tensor channel, we obtain the ground state mass to be 2.363(39) GeV and 2.384(67) GeV at $m_\pi\sim 938$ MeV and $650$ MeV, respectively.
In the pseudoscalar channel, using the gluonic operator whose continuum limit has the form of $\epsilon_{ijk}TrB_iD_jB_k$, the ground state mass is found to be 2.573(55) GeV and 2.585(65) GeV at the two pion masses. The masses of the tensor and pseudoscalar
glueballs do not show strong sea quark mass dependence in our study. However, since our pion masses are still
heavy, no decisive conclusions can be drawn on the quark mass dependence of glueball masses at present. In the scalar channel, the ground state masses extracted from the correlation functions of gluonic operators are determined to be around 1.4-1.5 GeV, which is close to the ground state masses from the correlation functions of the quark bilinear operators. One possible reason is the mixing between glueball states and conventional flavor singlet mesons, which requires further investigation in the future.

We also investigate the pseudoscalar channel using the topological charge density as the interpolation field operator, which is defined through Wilson loops and smeared by the Wilson flow technique. The masses of the lowest state derived in this way are much lighter (around 1 GeV) and compatible with the expected masses of the flavor singlet $q\bar{q}$ meson. This provides a strong hint that the operator $\epsilon_{ijk}TrB_iD_jB_k$ and the topological charge density (proportional to $Tr \mathbf{E}\cdot \mathbf{B}$) couple rather differently to the glueball states and $q\bar{q}$ mesons.

Admittedly the lattice volumes we used are relatively small and the continuum limit
remains to be taken, our current results are still helpful to clarify some aspects
of unquenched effects of glueballs and serves as a starting point for further studies.\\

\section*{Acknowledgements}
The numerical calculations are carried out on Tianhe-1A at the National Supercomputer Center (NSCC) in Tianjin and the GPU cluster
at Hunan Normal University. This work is supported in part by the National Science Foundation of China (NSFC) under Grants
No. 11575196, No. 11575197, No. 11335001, No. 11405053, No. 11405178 and No. 11275169. Y. C., Z. L. and C. L. also acknowledge the support of NSFC under
No. 11261130311 (CRC 110 by DFG and NSFC). Y. C. thanks the support by the CAS Center for Excellence in Particle Physics
(CCEPP). C.L. is also funded in part by National Basic Research Program of China (973 Program) under code number 2015CB856700.
M. G. thanks the support by the Youth Innovation Promotion Association of CAS (2015013).

\bibliography{ref}

\begin{thebibliography}{39}
\expandafter\ifx\csname natexlab\endcsname\relax\def\natexlab#1{#1}\fi
\expandafter\ifx\csname bibnamefont\endcsname\relax
  \def\bibnamefont#1{#1}\fi
\expandafter\ifx\csname bibfnamefont\endcsname\relax
  \def\bibfnamefont#1{#1}\fi
\expandafter\ifx\csname citenamefont\endcsname\relax
  \def\citenamefont#1{#1}\fi
\expandafter\ifx\csname url\endcsname\relax
  \def\url#1{\texttt{#1}}\fi
\expandafter\ifx\csname urlprefix\endcsname\relax\def\urlprefix{URL }\fi
\providecommand{\bibinfo}[2]{#2}
\providecommand{\eprint}[2][]{\url{#2}}

\bibitem[{\citenamefont{Jaffe and Johnson}(1976)}]{JAFFE1976201}
\bibinfo{author}{\bibfnamefont{R.}~\bibnamefont{Jaffe}} \bibnamefont{and}
  \bibinfo{author}{\bibfnamefont{K.}~\bibnamefont{Johnson}},
  \bibinfo{journal}{Physics Letters B} \textbf{\bibinfo{volume}{60}},
  \bibinfo{pages}{201 } (\bibinfo{year}{1976}).

\bibitem[{\citenamefont{Cornwall and Soni}(1983)}]{CORNWALL1983431}
\bibinfo{author}{\bibfnamefont{J.}~\bibnamefont{Cornwall}} \bibnamefont{and}
  \bibinfo{author}{\bibfnamefont{A.}~\bibnamefont{Soni}},
  \bibinfo{journal}{Physics Letters B} \textbf{\bibinfo{volume}{120}},
  \bibinfo{pages}{431 } (\bibinfo{year}{1983}).

\bibitem[{\citenamefont{Hou and Soni}(1984)}]{Hou:1982dy}
\bibinfo{author}{\bibfnamefont{W.-S.} \bibnamefont{Hou}} \bibnamefont{and}
  \bibinfo{author}{\bibfnamefont{A.}~\bibnamefont{Soni}},
  \bibinfo{journal}{Phys. Rev.} \textbf{\bibinfo{volume}{D29}},
  \bibinfo{pages}{101} (\bibinfo{year}{1984}).

\bibitem[{\citenamefont{Brower et~al.}(2000)\citenamefont{Brower, Mathur, and
  Tan}}]{Brower:2000rp}
\bibinfo{author}{\bibfnamefont{R.~C.} \bibnamefont{Brower}},
  \bibinfo{author}{\bibfnamefont{S.~D.} \bibnamefont{Mathur}},
  \bibnamefont{and} \bibinfo{author}{\bibfnamefont{C.-I.} \bibnamefont{Tan}},
  \bibinfo{journal}{Nucl. Phys.} \textbf{\bibinfo{volume}{B587}},
  \bibinfo{pages}{249} (\bibinfo{year}{2000}), \eprint{hep-th/0003115}.

\bibitem[{\citenamefont{Szczepaniak and Swanson}(2003)}]{Szczepaniak:2003mr}
\bibinfo{author}{\bibfnamefont{A.~P.} \bibnamefont{Szczepaniak}}
  \bibnamefont{and} \bibinfo{author}{\bibfnamefont{E.~S.}
  \bibnamefont{Swanson}}, \bibinfo{journal}{Phys. Lett.}
  \textbf{\bibinfo{volume}{B577}}, \bibinfo{pages}{61} (\bibinfo{year}{2003}),
  \eprint{hep-ph/0308268}.

\bibitem[{\citenamefont{Narison}(2006)}]{Narison:2005wc}
\bibinfo{author}{\bibfnamefont{S.}~\bibnamefont{Narison}},
  \bibinfo{journal}{Phys. Rev.} \textbf{\bibinfo{volume}{D73}},
  \bibinfo{pages}{114024} (\bibinfo{year}{2006}), \eprint{hep-ph/0512256}.

\bibitem[{\citenamefont{Sanchis-Alepuz
  et~al.}(2015)\citenamefont{Sanchis-Alepuz, Fischer, Kellermann, and von
  Smekal}}]{Sanchis-Alepuz:2015hma}
\bibinfo{author}{\bibfnamefont{H.}~\bibnamefont{Sanchis-Alepuz}},
  \bibinfo{author}{\bibfnamefont{C.~S.} \bibnamefont{Fischer}},
  \bibinfo{author}{\bibfnamefont{C.}~\bibnamefont{Kellermann}},
  \bibnamefont{and} \bibinfo{author}{\bibfnamefont{L.}~\bibnamefont{von
  Smekal}}, \bibinfo{journal}{Phys. Rev.} \textbf{\bibinfo{volume}{D92}},
  \bibinfo{pages}{034001} (\bibinfo{year}{2015}), \eprint{1503.06051}.

\bibitem[{\citenamefont{Klempt and Zaitsev}(2007)}]{Klempt:2007cp}
\bibinfo{author}{\bibfnamefont{E.}~\bibnamefont{Klempt}} \bibnamefont{and}
  \bibinfo{author}{\bibfnamefont{A.}~\bibnamefont{Zaitsev}},
  \bibinfo{journal}{Phys. Rept.} \textbf{\bibinfo{volume}{454}},
  \bibinfo{pages}{1} (\bibinfo{year}{2007}), \eprint{0708.4016}.

\bibitem[{\citenamefont{Mathieu et~al.}(2009)\citenamefont{Mathieu, Kochelev,
  and Vento}}]{Mathieu:2008me}
\bibinfo{author}{\bibfnamefont{V.}~\bibnamefont{Mathieu}},
  \bibinfo{author}{\bibfnamefont{N.}~\bibnamefont{Kochelev}}, \bibnamefont{and}
  \bibinfo{author}{\bibfnamefont{V.}~\bibnamefont{Vento}},
  \bibinfo{journal}{Int. J. Mod. Phys.} \textbf{\bibinfo{volume}{E18}},
  \bibinfo{pages}{1} (\bibinfo{year}{2009}), \eprint{0810.4453}.

\bibitem[{\citenamefont{Crede and Meyer}(2009)}]{Crede:2008vw}
\bibinfo{author}{\bibfnamefont{V.}~\bibnamefont{Crede}} \bibnamefont{and}
  \bibinfo{author}{\bibfnamefont{C.~A.} \bibnamefont{Meyer}},
  \bibinfo{journal}{Prog. Part. Nucl. Phys.} \textbf{\bibinfo{volume}{63}},
  \bibinfo{pages}{74} (\bibinfo{year}{2009}), \eprint{0812.0600}.

\bibitem[{\citenamefont{Ochs}(2013)}]{Ochs:2013gi}
\bibinfo{author}{\bibfnamefont{W.}~\bibnamefont{Ochs}}, \bibinfo{journal}{J.
  Phys.} \textbf{\bibinfo{volume}{G40}}, \bibinfo{pages}{043001}
  (\bibinfo{year}{2013}), \eprint{1301.5183}.

\bibitem[{\citenamefont{Morningstar and Peardon}(1997)}]{Morningstar:1997}
\bibinfo{author}{\bibfnamefont{C.~J.} \bibnamefont{Morningstar}}
  \bibnamefont{and} \bibinfo{author}{\bibfnamefont{M.}~\bibnamefont{Peardon}},
  \bibinfo{journal}{Phys. Rev. D} \textbf{\bibinfo{volume}{56}},
  \bibinfo{pages}{4043} (\bibinfo{year}{1997}).

\bibitem[{\citenamefont{Morningstar and Peardon}(1999)}]{Morningstar:1999rf}
\bibinfo{author}{\bibfnamefont{C.~J.} \bibnamefont{Morningstar}}
  \bibnamefont{and} \bibinfo{author}{\bibfnamefont{M.~J.}
  \bibnamefont{Peardon}}, \bibinfo{journal}{Phys. Rev.}
  \textbf{\bibinfo{volume}{D60}}, \bibinfo{pages}{034509}
  (\bibinfo{year}{1999}), \eprint{hep-lat/9901004}.

\bibitem[{\citenamefont{Chen et~al.}(2006)\citenamefont{Chen, Alexandru, Dong,
  Draper, Horv\'ath, Lee, Liu, Mathur, Morningstar, Peardon
  et~al.}}]{Chen:2005mg}
\bibinfo{author}{\bibfnamefont{Y.}~\bibnamefont{Chen}},
  \bibinfo{author}{\bibfnamefont{A.}~\bibnamefont{Alexandru}},
  \bibinfo{author}{\bibfnamefont{S.~J.} \bibnamefont{Dong}},
  \bibinfo{author}{\bibfnamefont{T.}~\bibnamefont{Draper}},
  \bibinfo{author}{\bibfnamefont{I.}~\bibnamefont{Horv\'ath}},
  \bibinfo{author}{\bibfnamefont{F.~X.} \bibnamefont{Lee}},
  \bibinfo{author}{\bibfnamefont{K.~F.} \bibnamefont{Liu}},
  \bibinfo{author}{\bibfnamefont{N.}~\bibnamefont{Mathur}},
  \bibinfo{author}{\bibfnamefont{C.}~\bibnamefont{Morningstar}},
  \bibinfo{author}{\bibfnamefont{M.}~\bibnamefont{Peardon}},
  \bibnamefont{et~al.}, \bibinfo{journal}{Phys. Rev. D}
  \textbf{\bibinfo{volume}{73}}, \bibinfo{pages}{014516}
  (\bibinfo{year}{2006}), \eprint{hep-lat/0510074}.

\bibitem[{\citenamefont{Gui et~al.}(2013)\citenamefont{Gui, Chen, Li, Liu, Liu,
  Ma, Yang, and Zhang}}]{Gui:2012gx}
\bibinfo{author}{\bibfnamefont{L.-C.} \bibnamefont{Gui}},
  \bibinfo{author}{\bibfnamefont{Y.}~\bibnamefont{Chen}},
  \bibinfo{author}{\bibfnamefont{G.}~\bibnamefont{Li}},
  \bibinfo{author}{\bibfnamefont{C.}~\bibnamefont{Liu}},
  \bibinfo{author}{\bibfnamefont{Y.-B.} \bibnamefont{Liu}},
  \bibinfo{author}{\bibfnamefont{J.-P.} \bibnamefont{Ma}},
  \bibinfo{author}{\bibfnamefont{Y.-B.} \bibnamefont{Yang}}, \bibnamefont{and}
  \bibinfo{author}{\bibfnamefont{J.-B.} \bibnamefont{Zhang}}
  (\bibinfo{collaboration}{CLQCD Collaboration}), \bibinfo{journal}{Phys. Rev.
  Lett.} \textbf{\bibinfo{volume}{110}}, \bibinfo{pages}{021601}
  (\bibinfo{year}{2013}), \eprint{1206.0125}.

\bibitem[{\citenamefont{Yang et~al.}(2013)\citenamefont{Yang, Gui, Chen, Liu,
  Liu, Ma, and Zhang}}]{Yang:2013xba}
\bibinfo{author}{\bibfnamefont{Y.-B.} \bibnamefont{Yang}},
  \bibinfo{author}{\bibfnamefont{L.-C.} \bibnamefont{Gui}},
  \bibinfo{author}{\bibfnamefont{Y.}~\bibnamefont{Chen}},
  \bibinfo{author}{\bibfnamefont{C.}~\bibnamefont{Liu}},
  \bibinfo{author}{\bibfnamefont{Y.-B.} \bibnamefont{Liu}},
  \bibinfo{author}{\bibfnamefont{J.-P.} \bibnamefont{Ma}}, \bibnamefont{and}
  \bibinfo{author}{\bibfnamefont{J.-B.} \bibnamefont{Zhang}}
  (\bibinfo{collaboration}{CLQCD Collaboration}), \bibinfo{journal}{Phys. Rev.
  Lett.} \textbf{\bibinfo{volume}{111}}, \bibinfo{pages}{091601}
  (\bibinfo{year}{2013}), \eprint{1304.3807}.

\bibitem[{\citenamefont{Ablikim et~al.}(2013)}]{Ablikim:2013hq}
\bibinfo{author}{\bibfnamefont{M.}~\bibnamefont{Ablikim}} \bibnamefont{et~al.}
  (\bibinfo{collaboration}{BESIII}), \bibinfo{journal}{Phys. Rev.}
  \textbf{\bibinfo{volume}{D87}}, \bibinfo{pages}{092009}
  (\bibinfo{year}{2013}), \bibinfo{note}{[Erratum: Phys.
  Rev.D87,no.11,119901(2013)]}, \eprint{1301.0053}.

\bibitem[{\citenamefont{Ablikim et~al.}(2016)}]{Ablikim:2016hlu}
\bibinfo{author}{\bibfnamefont{M.}~\bibnamefont{Ablikim}} \bibnamefont{et~al.}
  (\bibinfo{collaboration}{BESIII}), \bibinfo{journal}{Phys. Rev.}
  \textbf{\bibinfo{volume}{D93}}, \bibinfo{pages}{112011}
  (\bibinfo{year}{2016}), \eprint{1602.01523}.

\bibitem[{\citenamefont{Bali et~al.}(2000)\citenamefont{Bali, Bolder, Eicker,
  Lippert, Orth, Ueberholz, Schilling, and Struckmann}}]{Bali:2000vr}
\bibinfo{author}{\bibfnamefont{G.~S.} \bibnamefont{Bali}},
  \bibinfo{author}{\bibfnamefont{B.}~\bibnamefont{Bolder}},
  \bibinfo{author}{\bibfnamefont{N.}~\bibnamefont{Eicker}},
  \bibinfo{author}{\bibfnamefont{T.}~\bibnamefont{Lippert}},
  \bibinfo{author}{\bibfnamefont{B.}~\bibnamefont{Orth}},
  \bibinfo{author}{\bibfnamefont{P.}~\bibnamefont{Ueberholz}},
  \bibinfo{author}{\bibfnamefont{K.}~\bibnamefont{Schilling}},
  \bibnamefont{and}
  \bibinfo{author}{\bibfnamefont{T.}~\bibnamefont{Struckmann}}
  (\bibinfo{collaboration}{SESAM and T\ensuremath{\chi}L Collaborations}),
  \bibinfo{journal}{Phys. Rev. D} \textbf{\bibinfo{volume}{62}},
  \bibinfo{pages}{054503} (\bibinfo{year}{2000}), \eprint{hep-lat/0003012}.

\bibitem[{\citenamefont{Hart and Teper}(2002)}]{Hart:2001fp}
\bibinfo{author}{\bibfnamefont{A.}~\bibnamefont{Hart}} \bibnamefont{and}
  \bibinfo{author}{\bibfnamefont{M.}~\bibnamefont{Teper}}
  (\bibinfo{collaboration}{UKQCD Collaboration}), \bibinfo{journal}{Phys. Rev.
  D} \textbf{\bibinfo{volume}{65}}, \bibinfo{pages}{034502}
  (\bibinfo{year}{2002}), \eprint{hep-lat/0108022}.

\bibitem[{\citenamefont{Richards et~al.}(2010)\citenamefont{Richards, Irving,
  Gregory, and McNeile}}]{Richards:2010ck}
\bibinfo{author}{\bibfnamefont{C.~M.} \bibnamefont{Richards}},
  \bibinfo{author}{\bibfnamefont{A.~C.} \bibnamefont{Irving}},
  \bibinfo{author}{\bibfnamefont{E.~B.} \bibnamefont{Gregory}},
  \bibnamefont{and} \bibinfo{author}{\bibfnamefont{C.}~\bibnamefont{McNeile}}
  (\bibinfo{collaboration}{UKQCD Collaboration}), \bibinfo{journal}{Phys. Rev.
  D} \textbf{\bibinfo{volume}{82}}, \bibinfo{pages}{034501}
  (\bibinfo{year}{2010}), \eprint{1005.2473}.

\bibitem[{\citenamefont{Gregory et~al.}(2012)\citenamefont{Gregory, Irving,
  Lucini, McNeile, Rago, Richards, and Rinaldi}}]{Gregory:2012hu}
\bibinfo{author}{\bibfnamefont{E.}~\bibnamefont{Gregory}},
  \bibinfo{author}{\bibfnamefont{A.}~\bibnamefont{Irving}},
  \bibinfo{author}{\bibfnamefont{B.}~\bibnamefont{Lucini}},
  \bibinfo{author}{\bibfnamefont{C.}~\bibnamefont{McNeile}},
  \bibinfo{author}{\bibfnamefont{A.}~\bibnamefont{Rago}},
  \bibinfo{author}{\bibfnamefont{C.}~\bibnamefont{Richards}}, \bibnamefont{and}
  \bibinfo{author}{\bibfnamefont{E.}~\bibnamefont{Rinaldi}},
  \bibinfo{journal}{JHEP} \textbf{\bibinfo{volume}{10}}, \bibinfo{pages}{170}
  (\bibinfo{year}{2012}), \eprint{1208.1858}.

\bibitem[{\citenamefont{Fukaya et~al.}(2015)\citenamefont{Fukaya, Aoki, Cossu,
  Hashimoto, Kaneko, and Noaki}}]{Fukaya:2015ara}
\bibinfo{author}{\bibfnamefont{H.}~\bibnamefont{Fukaya}},
  \bibinfo{author}{\bibfnamefont{S.}~\bibnamefont{Aoki}},
  \bibinfo{author}{\bibfnamefont{G.}~\bibnamefont{Cossu}},
  \bibinfo{author}{\bibfnamefont{S.}~\bibnamefont{Hashimoto}},
  \bibinfo{author}{\bibfnamefont{T.}~\bibnamefont{Kaneko}}, \bibnamefont{and}
  \bibinfo{author}{\bibfnamefont{J.}~\bibnamefont{Noaki}}
  (\bibinfo{collaboration}{JLQCD Collaboration}), \bibinfo{journal}{Phys. Rev.
  D} \textbf{\bibinfo{volume}{92}}, \bibinfo{pages}{111501}
  (\bibinfo{year}{2015}), \eprint{1509.00944}.

\bibitem[{\citenamefont{Chowdhury et~al.}(2015)\citenamefont{Chowdhury,
  Harindranath, and Maiti}}]{Chowdhury:2014mra}
\bibinfo{author}{\bibfnamefont{A.}~\bibnamefont{Chowdhury}},
  \bibinfo{author}{\bibfnamefont{A.}~\bibnamefont{Harindranath}},
  \bibnamefont{and} \bibinfo{author}{\bibfnamefont{J.}~\bibnamefont{Maiti}},
  \bibinfo{journal}{Phys. Rev.} \textbf{\bibinfo{volume}{D91}},
  \bibinfo{pages}{074507} (\bibinfo{year}{2015}), \eprint{1409.6459}.

\bibitem[{\citenamefont{Su et~al.}(2006)\citenamefont{Su, Liu, Li, and
  Liu}}]{Su:2004sc}
\bibinfo{author}{\bibfnamefont{S.-q.} \bibnamefont{Su}},
  \bibinfo{author}{\bibfnamefont{L.-m.} \bibnamefont{Liu}},
  \bibinfo{author}{\bibfnamefont{X.}~\bibnamefont{Li}}, \bibnamefont{and}
  \bibinfo{author}{\bibfnamefont{C.}~\bibnamefont{Liu}}, \bibinfo{journal}{Int.
  J. Mod. Phys.} \textbf{\bibinfo{volume}{A21}}, \bibinfo{pages}{1015}
  (\bibinfo{year}{2006}), \eprint{hep-lat/0412034}.

\bibitem[{\citenamefont{Lepage and Mackenzie}(1993)}]{PhysRevD.48.2250}
\bibinfo{author}{\bibfnamefont{G.~P.} \bibnamefont{Lepage}} \bibnamefont{and}
  \bibinfo{author}{\bibfnamefont{P.~B.} \bibnamefont{Mackenzie}},
  \bibinfo{journal}{Phys. Rev. D} \textbf{\bibinfo{volume}{48}},
  \bibinfo{pages}{2250} (\bibinfo{year}{1993}).

\bibitem[{\citenamefont{Umeda et~al.}(2003)\citenamefont{Umeda, Aoki, Fukugita,
  Ishikawa, Ishizuka, Iwasaki, Kanaya, Kuramashi, Lesk, Namekawa
  et~al.}}]{Umeda:2003pj}
\bibinfo{author}{\bibfnamefont{T.}~\bibnamefont{Umeda}},
  \bibinfo{author}{\bibfnamefont{S.}~\bibnamefont{Aoki}},
  \bibinfo{author}{\bibfnamefont{M.}~\bibnamefont{Fukugita}},
  \bibinfo{author}{\bibfnamefont{K.-I.} \bibnamefont{Ishikawa}},
  \bibinfo{author}{\bibfnamefont{N.}~\bibnamefont{Ishizuka}},
  \bibinfo{author}{\bibfnamefont{Y.}~\bibnamefont{Iwasaki}},
  \bibinfo{author}{\bibfnamefont{K.}~\bibnamefont{Kanaya}},
  \bibinfo{author}{\bibfnamefont{Y.}~\bibnamefont{Kuramashi}},
  \bibinfo{author}{\bibfnamefont{V.~I.} \bibnamefont{Lesk}},
  \bibinfo{author}{\bibfnamefont{Y.}~\bibnamefont{Namekawa}},
  \bibnamefont{et~al.} (\bibinfo{collaboration}{CP-PACS Collaboration}),
  \bibinfo{journal}{Phys. Rev. D} \textbf{\bibinfo{volume}{68}},
  \bibinfo{pages}{034503} (\bibinfo{year}{2003}), \eprint{hep-lat/0302024}.

\bibitem[{\citenamefont{Klassen}(1998)}]{Klassen:1998ua}
\bibinfo{author}{\bibfnamefont{T.~R.} \bibnamefont{Klassen}},
  \bibinfo{journal}{Nucl. Phys.} \textbf{\bibinfo{volume}{B533}},
  \bibinfo{pages}{557} (\bibinfo{year}{1998}), \eprint{hep-lat/9803010}.

\bibitem[{\citenamefont{Edwards et~al.}(2008)\citenamefont{Edwards, Joo, and
  Lin}}]{Edwards:2008ja}
\bibinfo{author}{\bibfnamefont{R.~G.} \bibnamefont{Edwards}},
  \bibinfo{author}{\bibfnamefont{B.}~\bibnamefont{Joo}}, \bibnamefont{and}
  \bibinfo{author}{\bibfnamefont{H.-W.} \bibnamefont{Lin}},
  \bibinfo{journal}{Phys. Rev.} \textbf{\bibinfo{volume}{D78}},
  \bibinfo{pages}{054501} (\bibinfo{year}{2008}), \eprint{0803.3960}.

\bibitem[{\citenamefont{Shuryak and Verbaarschot}(1995)}]{PhysRevD.52.295}
\bibinfo{author}{\bibfnamefont{E.~V.} \bibnamefont{Shuryak}} \bibnamefont{and}
  \bibinfo{author}{\bibfnamefont{J.~J.~M.} \bibnamefont{Verbaarschot}},
  \bibinfo{journal}{Phys. Rev. D} \textbf{\bibinfo{volume}{52}},
  \bibinfo{pages}{295} (\bibinfo{year}{1995}).

\bibitem[{\citenamefont{Moran and Leinweber}(2008)}]{Moran:2008qd}
\bibinfo{author}{\bibfnamefont{P.~J.} \bibnamefont{Moran}} \bibnamefont{and}
  \bibinfo{author}{\bibfnamefont{D.~B.} \bibnamefont{Leinweber}},
  \bibinfo{journal}{Phys. Rev.} \textbf{\bibinfo{volume}{D78}},
  \bibinfo{pages}{054506} (\bibinfo{year}{2008}), \eprint{0801.2016}.

\bibitem[{\citenamefont{L\"uscher}(2010)}]{Luscher:2010iy}
\bibinfo{author}{\bibfnamefont{M.}~\bibnamefont{L\"uscher}},
  \bibinfo{journal}{JHEP} \textbf{\bibinfo{volume}{08}}, \bibinfo{pages}{071}
  (\bibinfo{year}{2010}), \bibinfo{note}{[Erratum: JHEP03,092(2014)]},
  \eprint{1006.4518}.

\bibitem[{\citenamefont{Borsanyi et~al.}(2012)}]{Borsanyi:2012zs}
\bibinfo{author}{\bibfnamefont{S.}~\bibnamefont{Borsanyi}}
  \bibnamefont{et~al.}, \bibinfo{journal}{JHEP} \textbf{\bibinfo{volume}{09}},
  \bibinfo{pages}{010} (\bibinfo{year}{2012}), \eprint{1203.4469}.

\bibitem[{\citenamefont{Witten}(1979)}]{Witten:1979vv}
\bibinfo{author}{\bibfnamefont{E.}~\bibnamefont{Witten}},
  \bibinfo{journal}{Nucl. Phys.} \textbf{\bibinfo{volume}{B156}},
  \bibinfo{pages}{269} (\bibinfo{year}{1979}).

\bibitem[{\citenamefont{Veneziano}(1979)}]{VENEZIANO1979213}
\bibinfo{author}{\bibfnamefont{G.}~\bibnamefont{Veneziano}},
  \bibinfo{journal}{Nuclear Physics B} \textbf{\bibinfo{volume}{159}},
  \bibinfo{pages}{213 } (\bibinfo{year}{1979}), ISSN \bibinfo{issn}{0550-3213}.

\bibitem[{\citenamefont{Simonov}(1989)}]{Simonov:1988xs}
\bibinfo{author}{\bibfnamefont{{\relax Yu}.~A.} \bibnamefont{Simonov}},
  \bibinfo{journal}{Sov. J. Nucl. Phys.} \textbf{\bibinfo{volume}{50}},
  \bibinfo{pages}{134} (\bibinfo{year}{1989}), \bibinfo{note}{[Yad.
  Fiz.50,213(1989)]}.

\bibitem[{\citenamefont{Helmes et~al.}(2017)\citenamefont{Helmes, Knippschild,
  Kostrzewa, Liu, Jost, Ottnad, Urbach, Wenger, and Werner}}]{Helmes:2017ccf}
\bibinfo{author}{\bibfnamefont{C.}~\bibnamefont{Helmes}},
  \bibinfo{author}{\bibfnamefont{B.}~\bibnamefont{Knippschild}},
  \bibinfo{author}{\bibfnamefont{B.}~\bibnamefont{Kostrzewa}},
  \bibinfo{author}{\bibfnamefont{L.}~\bibnamefont{Liu}},
  \bibinfo{author}{\bibfnamefont{C.}~\bibnamefont{Jost}},
  \bibinfo{author}{\bibfnamefont{K.}~\bibnamefont{Ottnad}},
  \bibinfo{author}{\bibfnamefont{C.}~\bibnamefont{Urbach}},
  \bibinfo{author}{\bibfnamefont{U.}~\bibnamefont{Wenger}}, \bibnamefont{and}
  \bibinfo{author}{\bibfnamefont{M.}~\bibnamefont{Werner}}
  (\bibinfo{collaboration}{ETM}) (\bibinfo{year}{2017}), \eprint{1710.03698}.

\bibitem[{\citenamefont{Christ et~al.}(2010)\citenamefont{Christ, Dawson,
  Izubuchi, Jung, Liu, Mawhinney, Sachrajda, Soni, and Zhou}}]{Christ:2010dd}
\bibinfo{author}{\bibfnamefont{N.~H.} \bibnamefont{Christ}},
  \bibinfo{author}{\bibfnamefont{C.}~\bibnamefont{Dawson}},
  \bibinfo{author}{\bibfnamefont{T.}~\bibnamefont{Izubuchi}},
  \bibinfo{author}{\bibfnamefont{C.}~\bibnamefont{Jung}},
  \bibinfo{author}{\bibfnamefont{Q.}~\bibnamefont{Liu}},
  \bibinfo{author}{\bibfnamefont{R.~D.} \bibnamefont{Mawhinney}},
  \bibinfo{author}{\bibfnamefont{C.~T.} \bibnamefont{Sachrajda}},
  \bibinfo{author}{\bibfnamefont{A.}~\bibnamefont{Soni}}, \bibnamefont{and}
  \bibinfo{author}{\bibfnamefont{R.}~\bibnamefont{Zhou}},
  \bibinfo{journal}{Phys. Rev. Lett.} \textbf{\bibinfo{volume}{105}},
  \bibinfo{pages}{241601} (\bibinfo{year}{2010}), \eprint{1002.2999}.

\bibitem[{\citenamefont{Michael et~al.}(2013)\citenamefont{Michael, Ottnad, and
  Urbach}}]{Michael:2013gka}
\bibinfo{author}{\bibfnamefont{C.}~\bibnamefont{Michael}},
  \bibinfo{author}{\bibfnamefont{K.}~\bibnamefont{Ottnad}}, \bibnamefont{and}
  \bibinfo{author}{\bibfnamefont{C.}~\bibnamefont{Urbach}}
  (\bibinfo{collaboration}{ETM}), \bibinfo{journal}{Phys. Rev. Lett.}
  \textbf{\bibinfo{volume}{111}}, \bibinfo{pages}{181602}
  (\bibinfo{year}{2013}), \eprint{1310.1207}.

\end{thebibliography}

\end{document}